\documentclass[journal=jacsat,manuscript=article]{achemso}
\usepackage{graphicx,epstopdf}
\usepackage[version=3]{mhchem} 

\author{Hossein Mirhosseini}
\affiliation{Dynamics of Condensed Matter and Center for Sustainable Systems Design, Chair of Theoretical Chemistry, University of Paderborn, 33098 Warburger Str. 100, Paderborn, Germany}
\email{mirhosse@mail.uni-paderborn.de}

\author{Hossein Tahmasbi}
\affiliation
{Leiden Institute of Chemistry, Gorlaeus Laboratories, Leiden University, P.O. Box 9502, 2300 RA Leiden, The Netherlands}
\author{Sai Ram Kuchana}
\affiliation{Dynamics of Condensed Matter and Center for Sustainable Systems Design, Chair of Theoretical Chemistry, University of Paderborn, 33098 Warburger Str. 100, Paderborn, Germany}
\author{S. Alireza Ghasemi}
\affiliation{Dynamics of Condensed Matter and Center for Sustainable Systems Design, Chair of Theoretical Chemistry, University of Paderborn, 33098 Warburger Str. 100, Paderborn, Germany}
\author{Thomas D. K\"uhne}
\affiliation{Dynamics of Condensed Matter and Center for Sustainable Systems Design, Chair of Theoretical Chemistry, University of Paderborn, 33098 Warburger Str. 100, Paderborn, Germany}
\email{tdkuehne@mail.uni-paderborn.de}
\title[]
  {An automated approach for developing neural network interatomic potentials with FLAME}


\begin{document}


\begin{abstract}
The performance of machine learning interatomic potentials relies on the quality of the training dataset. In this work, we present an approach for generating diverse and representative training data points which initiates with \emph{ab initio} calculations for bulk structures. The data generation and potential construction further proceed side-by-side in a cyclic process of training the neural network and crystal structure prediction based on the developed interatomic potentials. All steps of the data generation and potential development are performed with minimal human intervention. We show the reliability of our approach by assessing the performance of neural network potentials developed for two inorganic systems. 
\end{abstract}


\section{Introduction}
In the context of molecular dynamics (MD) simulation, the classical force field and \emph{ab initio} MD are the two limits. While the former is known to be more efficient and the method of choice for large-scale simulations, the latter has proved to be a more accurate method and is applicable for the short-scale simulation of systems with less than a few hundred atoms.
Atomistic modeling based on machine learning interatomic potentials (MLIPs), on the other hand, has the advantages of both methods. Meaning, MLIPs can potentially enable performance of large-scale simulations (large length scales and long time scales) with quantum accuracy orders of magnitude faster than current methods based on density functional theory (DFT). 

The success of MD simulations relies on the accuracy of the potential energy surface (PES). In the context of MLIPs, the PES is described as a function of atomic coordinates through local environment descriptors~\cite{Behler2011}.
MLIPs employ machine learning (ML) algorithms and local environment descriptors to describe atomic interactions. Numerous MLIPs have emerged in recent years. Examples of such potentials include high-dimensional neural network potentials (NNPs) ~\cite{Behler2007} and Gaussian approximation potential (GAP) ~\cite{Bartok2010}, among others~\cite{Botu2014,Thompson2015,Li2015,Hansen2015,Brockherde2017,Smith2017,Yao2018,Podryabinkin2017,Zhang2018,Bereau2018,Schuett2019}. The remarkable performance of MLIPs has already been demonstrated for inorganic solids~\cite{Behler2007,PhysRevB.81.100103,PhysRevB.81.184107,Artrith2011,Khaliullin2011,PhysRevLett.108.115701,Artrith2016}, hybrid materials~\cite{Eckhoff2019}, water~\cite{Morawietz2016,Sukuba2018}, 
interfaces~\cite{Natarajan2016,Quaranta2017,Quaranta2019,Hellstroem2019,Ludwig2019}, and the dynamics of defects in crystalline and amorphous materials~\cite{Li2017,Korolev2020,Elbaz2020,Xu2020}. 

A key component for developing reliable potentials in ML-based PES models is building a robust and representative reference dataset for model training. The training dataset is usually composed of atomic configurations with corresponding energies, forces, and/or stress tensors from DFT calculations. 
A diverse and extensive dataset consists of not only data points for configurations close to equilibrium but also those for energetically less favorable configurations to avoid the tendency of developing biased potentials and overfitted models. Often, training datasets are generated by performing MD calculations at various temperatures and/or densities, followed by electronic structure calculations for snapshots extracted from MD trajectories. \emph{Ab initio} MD calculations for large systems, however, are demanding, making the development of MLIPs slow and computationally expensive.

In this work, we adopt a different approach to generating robust and representative training datasets. Namely, we perform \emph{ab initio} geometry optimization solely for periodic bulk structures and extract samples from geometry optimization calculations to initiate training of neural network (NN) interatomic potentials.  The bulk structures are taken from the Materials Project database (MPDB)~\cite{MPD1,MPD2}. In addition, crystal structure prediction based on ionic substitution~\cite{Hautier2011} is performed to increase the diversity of bulk structures.
NN interatomic potentials are further trained in iterative training cycles by including data from crystal structure prediction based on the minima hopping method~\cite{Goedecker2004,Amsler2010}. Although hops from one local minimum to another in minima hopping is based on MD calculations, these calculations are performed with generated NNPs, therefore, they are not computationally demanding. 

Generating extensive training datasets requires performing a substantial number of calculations and handling a large amount of data. An important task in modern computational materials science is to accomplish these tasks automatically. In this respect, our Python-based script generates NNPs with minimal human intervention. To demonstrate the performance of NNPs constructed with our approach, we compare the results of molecular statics and molecular dynamics calculations using the NNPs to reference DFT results. 

\section{Computational details}
All DFT calculations were carried out by the Vienna Ab initio Simulation Package (VASP)~\cite{Kresse1993,Kresse1996,Kresse1996a} employing projector augmented  wave (PAW) pseudopotentials~\cite{bloechl_1994}. A plane-wave cutoff  of $520$~eV and the Perdew-Burke-Ernzerhof~\cite{Perdew1996} (PBE) form of the generalized gradient approximation for the exchange-correlation potential were used for geometry optimization and single point energy calculations.

Phonon dispersion relations were calculated using the frozen phonon approach as implemented in the PHONOPY package~\cite{Togo2015}. Convergence tests were performed to obtain converged force constants and phonon density of states. For calculating lattice thermal conductivity, we used the ShengBTE code~\cite{Li2014Jun}, which solves the Boltzmann transport equation for phonons.
This code requires the second- and third-order interatomic force constants (IFCs), which were computed with PHONOPY and thirdorder.py~\cite{Li2014Jun}, respectively. The second- and third-order IFCs were calculated with $3\times3\times3$ (for CuInSe$_2$) and $4\times4\times4$ (for the anatase phase of TiO$_2$) supercells. For the third-order IFCs, the interactions up to eleventh nearest neighbors were considered, for which the lattice thermal conductivity of both structures are converged.

To minimize human control in the training process, we took advantage of atomate~\cite{atomate}, an open-source Python framework for materials science simulation and analysis with an emphasis on automation. Atomate is built on top of open-source libraries such as pymatgen~\cite{pymatgen} (an open-source Python library for materials analysis), Custodian (a just-in-time job management framework), and FireWorks~\cite{Jain2015} (a code for defining, managing, and executing workflows), allowing automatic job submission and result collection. Atomic configurations are taken from the MPDB, which makes its data available  through the open Materials Application Programming Interface (API) and the pymatgen materials analysis package. 

We trained and utilized high-dimensional NNPs using algorithms implemented in FLAME~\cite{Amsler2020}, an open-source software package~\cite{flame} that enables performance of a wide range of atomistic simulations for the construction and exploration of PESs. FLAME has been used in several development and application studies in the context of MLIPs~\cite{Ghasemi2015,Rostami2018,Faraji2017,Rasoulkhani2017,Asna2017,Faraji2019}. We refer to Reference~\cite{Amsler2020} for more details. 

In our approach, two crystal structure prediction methods are employed to increase the diversity of the training dataset: crystal structure prediction based on  minima  hopping~\cite{Goedecker2004,Amsler2010} and that on ionic substitutions~\cite{Hautier2011}. Minima hopping is a method for finding the global minimum of a PES. It not only searches for the global minimum but also explores low-energy configurations. 
Ionic-substitution crystal structure prediction is based on the common approach to proposing new compounds by replacing one ion with another chemically similar ion. The mathematical model provides a probability distribution for ionic substitutions that it has learned from a database of experimentally observed crystal structures.

To compare atomic configurations generated during the potential development process, the distances between structural fingerprints~\cite{Oganov2009, Behler2011} are calculated. If the distance between two fingerprints is less than a defined value, then the structures are considered similar. 

\section{Workflow}
\begin{figure*}[ht]
	\centering
\includegraphics[scale=.60]{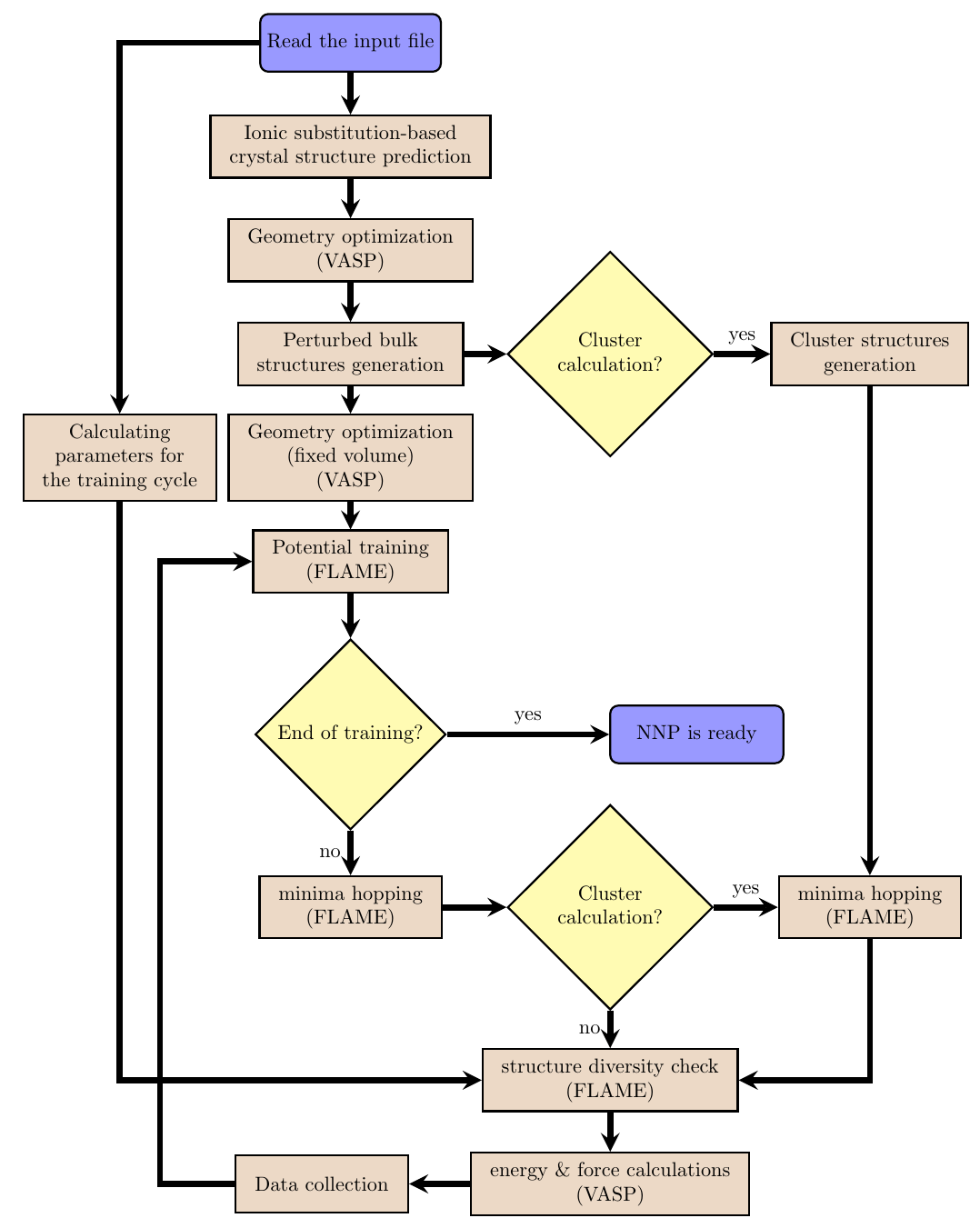}
	\caption{The workflow diagram of the automated algorithm to generate reference datasets and train neural network interatomic potentials.}
	\label{FIG:flowchart}
\end{figure*}

In the following sections, we describe the workflow of our script, which is shown in Figure~\ref{FIG:flowchart}. In the course of constructing NNPs, the script keeps track of its steps. If a failure occurs and the script cannot advance, the user can restart the script from the last successfully accomplished step. It is noted that the user can always restart the script from a previous step with different parameters.

\subsection{Input data}
The input file is in YAML format~\cite{YAML}. The composition of the system is determined by providing either a) a Materials Project ID (mp-id), b) a species (element symbols and their oxidation states, e.g. (Ag, +1)), or c) a single element symbol. In the last case, only structures with the given element available in the MPDB will be considered. If the user provides an mp-id or species, then all compounds containing the specified elements with the given oxidation states are considered.  For example, CuInSe$_{\rm 2}$ and CuIn$_{5}$Se$_{\rm 8}$ will be included in the training process for the (In, +3)-(Cu, +1)-(Se, -2) system, because the oxidation states of the constituent elements are the same in both compounds. It should be noted that the transferability of NNPs with no restriction on the stoichiometry of training systems needs a careful examination~\cite{Mangold2020,Benoit2021}.
There is an option to limit training data points to systems with a fixed stoichiometry.

The size of the system is established by specifying the maximum number of atoms in bulk structures. Structures containing less than 4 atoms are discarded.
If the training dataset should include clusters, then the minimum and maximum number of atoms in clusters should be specified in the input file.

Parameters needed for training cycles, such as the number of nodes in hidden layers of the NN and temperatures for performing minima hopping, can be specified in the input file. Parameters that are needed for structure diversity check, such as the minimum allowed distance between atoms and the average of distances between the fingerprints of sample structures, are calculated based on the given composition. We discuss these parameters in the corresponding sections.

\subsection{Crystal structure prediction}
The next step after selecting the system is crystal structure prediction based on ionic substitution~\cite{Hautier2011}. The aim of this step is to include as many bulk structures as possible in the data generation process. After new bulk structures are found, the duplicated structures are removed from the final list and the remaining structures are optimized until the residual force on each atom is less than 0.05 eV/\AA. 

In addition to ionic substitution for crystal structure prediction, the user can substitute a fraction of atoms to include more elements in the system. The purpose of this ion substitution is to extend the chemical space to systems that do not exist in databases. For example, a given percentage of In in the In-Cu-Se system can be replaced with Tl to build the quaternary system of In-Tl-Cu-Se.

\subsection{Perturbed structures creation}
To expand the reference training dataset, periodic bulk structures are perturbed. That is, atoms are randomly displaced from their optimized positions by translating sites along a vector and/or rotating sites by an angle around a vector. In addition, stressed structures (expanded/contracted bulk structures) are generated. All these structures are loosely optimized, and samples are taken
from the different steps of the geometry optimization calculations.
The structure relaxation stops when the residual forces on the atoms
are less than $0.5$~eV/\AA. Up to four samples are taken from each geometry optimization calculation: when the maximum total force on atoms is between $2$ and $3$~eV/\AA, between $1$ and $2$~eV/\AA, and between $0.5$ and $1$~eV/\AA. One sample is taken from the optimized structure.

\subsection{Training cycle}
The training cycle has five steps: potential training, minima hopping, structure diversity check, single point energy calculations, and data collection. The first three steps are performed as implemented in FLAME. For training NNPs, the data obtained from the previous step is divided into $90$\% for training and $10$\% for validation. The generated NNP in each cycle are used for performing minima hopping. Seed configurations for minima hopping are the bulk structures from the crystal structure prediction step. 

In the diversity check step, the snapshots taken from minima hopping trajectories are screened to remove similar and nonphysical structures (due to the small distance of atoms or an unacceptable density). To remove similar structures, the distances between all structural fingerprints of atomic configurations are computed. If the distance between a pair of fingerprints is smaller than the accepted value, then one of the configurations is discarded. It is noted that parameters for structure diversity check are adjusted for each cycle of training. That is, the allowed minimum distance between atoms in the earlier cycles of training (when the model is not yet well-trained) is larger than that in the later steps of training (when NNPs are more reliable). In contrast, while during the early cycles of training, the tolerance for identifying similar structures is large (loose), the tolerance becomes smaller in the later cycles of training to select only structures with a large distance from the previously selected structures.

The structures that pass diversity check are sent for single point energy calculations at the DFT level. In the data collection step, the forces and energies calculated by VASP are collected to be used in the next cycle of training. Configurations for which the residual force on their atoms is less than $0.10$~eV/\AA~ are stored for minima hopping in the next cycle. The training cycle is repeated until the desired number of training cycles, specified in the input file, is reached.

\section{Validation of the approach}
In the following sections, we demonstrate the validity of our approach for developing interatomic potentials by comparing DFT and NNP results for the (Ti, +4)-(O, -2) and (In, +3)-(Cu, +1)-(Se, -2) systems.

\subsection{The (Ti, +4)-(O, -2) system}
We have developed a NNP for the (Ti, +4)-(O, -2) system with a fixed stoichiometry that limits the composition to titanium dioxide (TiO$_{\rm 2}$). The diverse structures of TiO$_{\rm 2}$ makes it an ideal benchmark material to evaluate the performance of the generated NNP. Among the many applications of TiO$_{\rm 2}$, the ability of TiO$_{\rm 2}$ to photocatalyze water splitting reactions makes it interesting for sustainable energy applications~\cite{Fujishima1972,Ma2014,Ni2007}.

For developing (Ti, +4)-(O, -2) potentials, $4606$ VASP geometry optimization calculations were performed for $223$ dissimilar structures. Of these structures, $77$ were taken from the MPDB and the rest was found by crystal structure prediction. The training process started with $16696$ data points and stopped after five steps, as it was specified in the input file. The total number of training points after five cycles of training was $67248$.  

\begin{figure}[t]
\centering
\includegraphics[scale=.9]{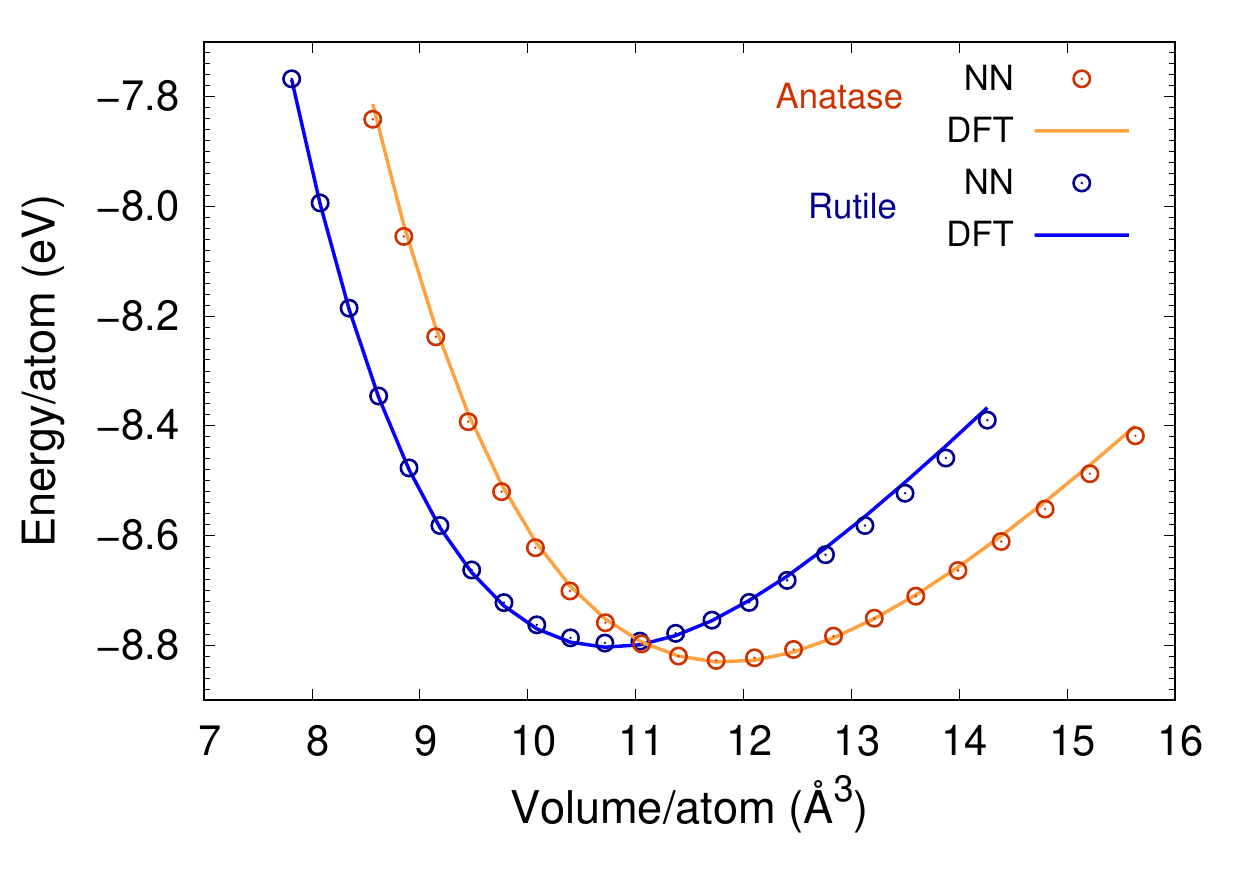}
\caption{Comparison of equation of state for anatase and rutile TiO$_{\rm 2}$ calculated by the NNP and DFT.}
\label{FIG:TiO2_EOS}
\end{figure}

\begin{figure}[ht]
\centering
\includegraphics[scale=.9]{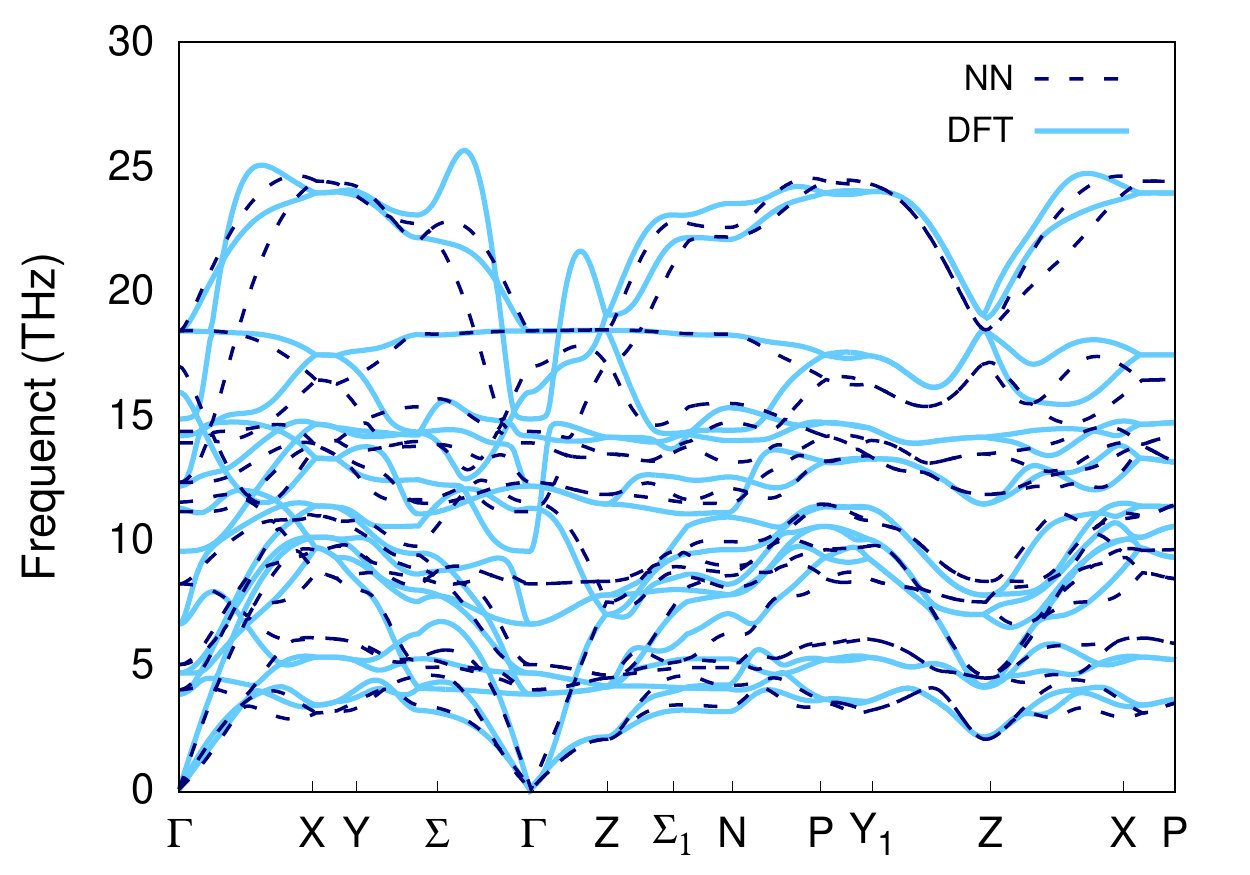}
\caption{Phonon dispersion relation of anatase TiO$_{\rm 2}$ calculated by DFT and the NNP.}
\label{FIG:TiO2_phonon}
\end{figure}

\begin{figure}[ht]
\centering
\includegraphics[scale=.9]{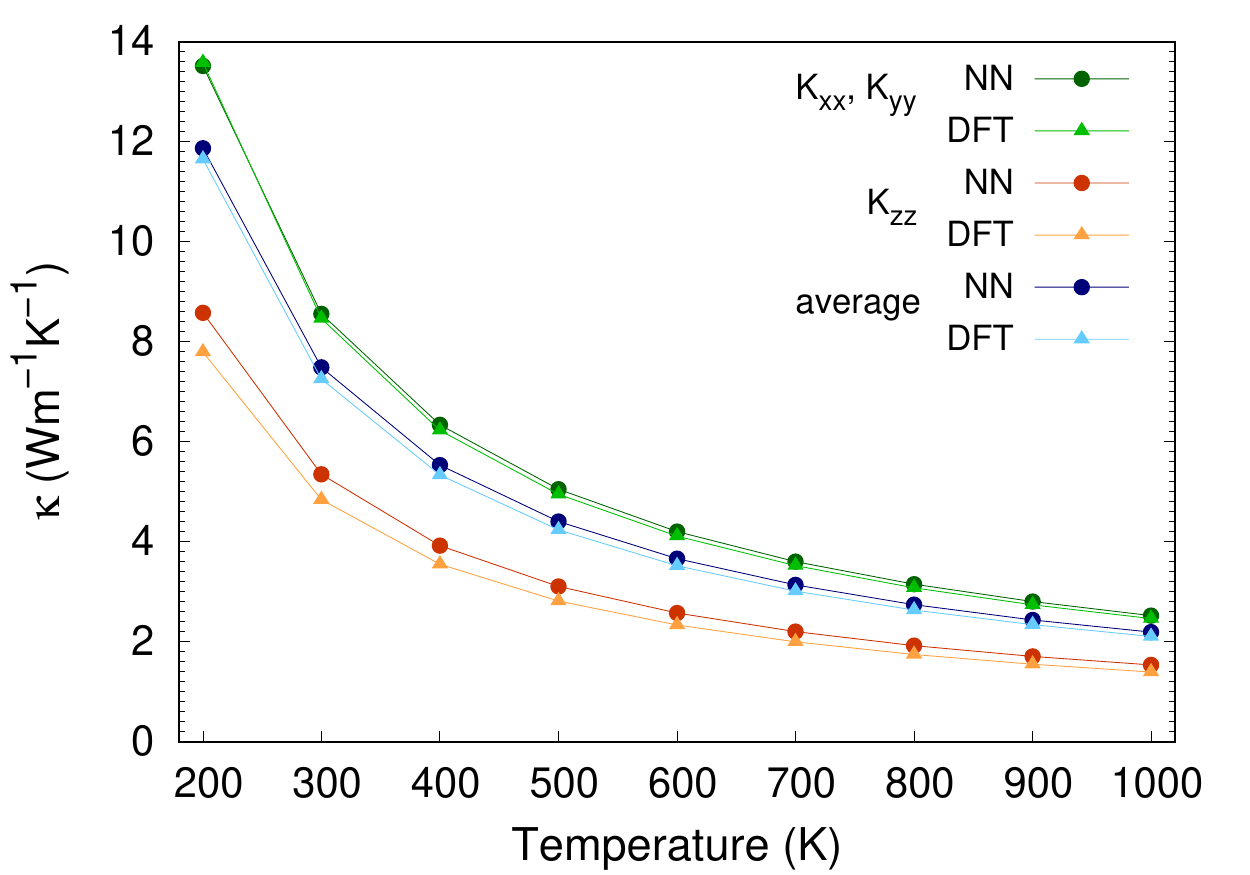}
\caption{Lattice thermal conductivity of the anatase phase of TiO$_{\rm 2}$ calculated with DFT and the NNP.}
\label{FIG:TiO2_kappa}
\end{figure}

To asses the performance of the developed potentials for the (Ti, +4)-(O, -2) system, first, the equation of states (EOS) for two phases of TiO$_{\rm 2}$ were calculated (see Figure~\ref{FIG:TiO2_EOS}). The agreement between \emph{ab initio} and NNP results is very good. The rutile phase of TiO$_{\rm 2}$ is more stable than the anatase phase. DFT calculations with PBE, however, predict anatase to be more stable than rutile~\cite{Asna2017}. The phonon dispersion relation calculated for a $4\times4\times4$ supercell of anatase TiO$_{\rm 2}$ is shown in Figure~\ref{FIG:TiO2_phonon}. The main contribution to lattice thermal conductivity is coming from acoustic branches~\cite{Slack1973}. In this frequency range, the agreement between DFT and NNP results is very good. Therefore, a good agreement between lattice thermal conductivity calculated by DFT and by the trained NNP is expected (see Figure~\ref{FIG:TiO2_kappa}). Phonon dispersion relations and lattice thermal conductivity values calculated for TiO$_{\rm 2}$ by different exchange-correlation functionals agree less with each other~\cite{Shojaee2009,Arrigoni2019,Torres2019}. For example, the lattice thermal conductivity of anatase calculated with the local density approximation functional at 300 K is 3.6 Wm$^{-1}$K$^{-1}$ larger than that calculated with PBE (7.2 Wm$^{-1}$K$^{-1}$). The difference between the PBE and NNP results is well below this value.

\begin{figure}[ht]
\centering
\includegraphics[scale=.9]{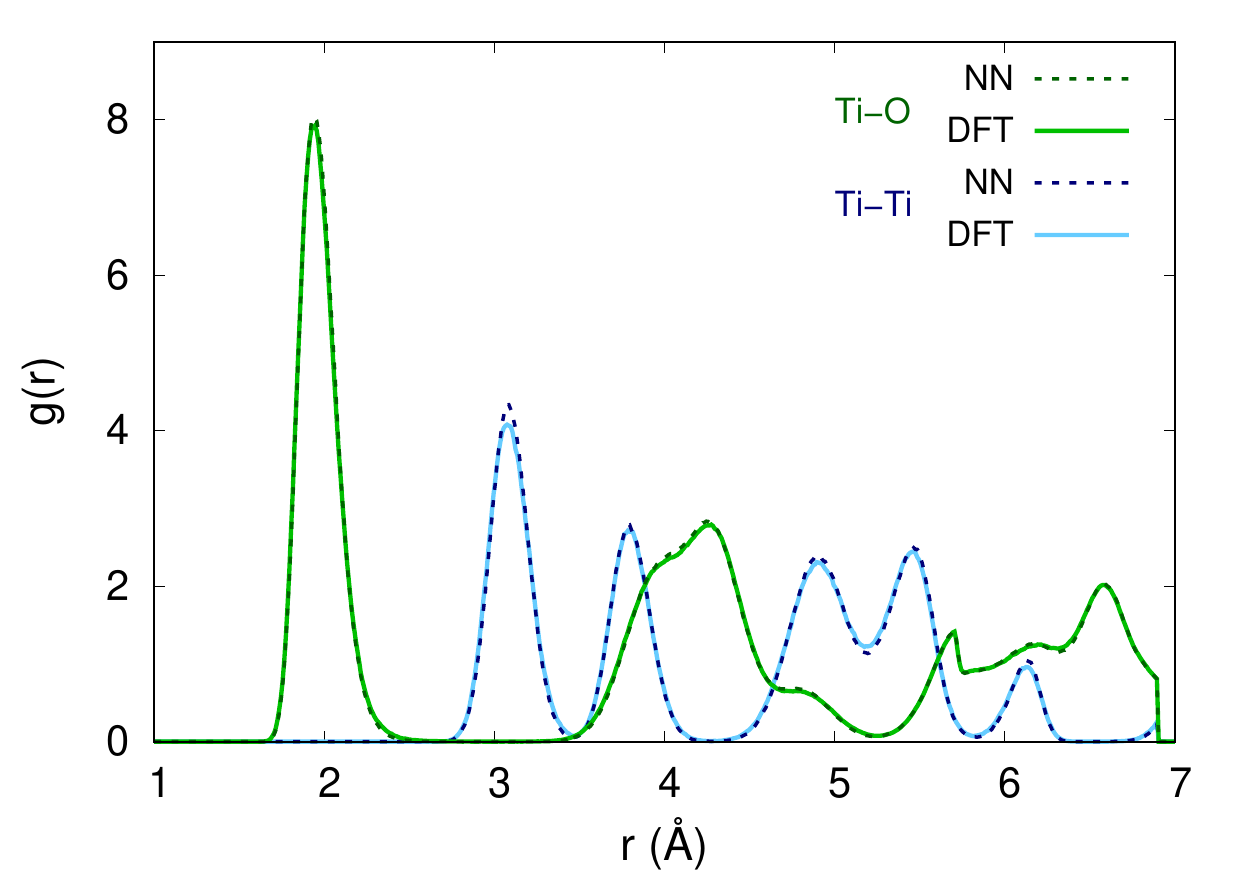}
\caption{The radial distribution functions for anatase TiO$_{\rm 2}$ computed with DFT and the NNP at T= 1000 K.}
\label{FIG:TiO2_VASP-vs-FLAME-1000K}
\end{figure}

To evaluate the accuracy of Ti-O potentials for MD simulations, NVT MD calculations at various temperatures were performed by employing MD algorithms implemented in EON~\cite{EON}. To perform MD calculations with NNPs, EON is coupled with FLAME. A comparison between radial distribution functions calculated by DFT and by the constructed NNP for a 108-atom supercell of anatase is shown in Figure~\ref{FIG:TiO2_VASP-vs-FLAME-1000K}. MD calculations were performed at $1000$~K for $35000$ MD steps after equilibration. The very good agreement between the DFT and NNP results shows the reliability of the developed NNP for MD simulations. 

As mentioned earlier, only periodic bulk structures have been used for training NN interatomic potentials. It is interesting to see whether the developed  potentials could produce reasonable results for non-periodic systems, like slabs. Therefore, we calculated the energy per atom for a 2D phase of TiO$_{\rm 2}$ that was recently discovered by crystal structure prediction~\cite{Asna2017}. For a single layer of this 2D structure with 12 atoms, the difference between the energy per atom calculated with DFT and that with the NNP is as small as 3.4 meV. It is worth noting that this structure was not in the training dataset. In addition, we have calculated the surface energies of the non-polar (001) surface of rutile and anatase. The surface energy for the (001) surface of rutile calculated by DFT and the NNP is 0.13 eV/\AA$^{2}$ (2.10 J/m$^{2}$) and 0.13 eV/\AA$^{2}$ (2.04 J/m$^{2}$), respectively. For the anatase phase, the surface energies were computed as 0.13 eV/\AA$^{2}$ (2.07 J/m$^{2}$) and 0.13 eV/\AA$^{2}$ (2.06 J/m$^{2}$) by DFT and the NNP, respectively. 

\subsection{The (In, +3)-(Cu, +1)-(Se, -2) system}
The representative of the (In, +3)-(Cu, +1)-(Se, -2) system is copper indium selenide (CuInSe$_{\rm 2}$), one of the most promising absorbers for thin-film solar cells~\cite{Frontier2019,Regmi2020}. The applications of CuInSe$_{\rm 2}$ and its alloy with Ga in conversion of solar energy into sustainable energy are not limited to photovoltaics~\cite{Kim2019,Hu2020}. Furthermore, in the last few years, we and many other groups have studied the characteristics of CuInSe$_{\rm 2}$ using first-principle methods~\cite{Mirhosseini2020}, so a large amount of \emph{ab initio} data is available for this system to benchmark against.

The process of developing the NNP started with finding structures that contain (In, +3), (Cu, +1), and (Se, -2). For this system, there are six known structures with less than 40 atoms in the MPDB. Two of these six structures have oxidation states that do not match the oxidation states of the species specified in the input file (In$_{\rm 2}$CuSe$_{\rm 4}$ and In$_{\rm 8}$Cu$_{\rm 7}$Se$_{\rm 16}$). In the crystal structure prediction step, $52$ dissimilar structures were found. The first cycle of training started with the data obtained from the optimization of perturbed/stressed structures.
In total, $1729$ VASP geometry optimization calculations for bulk structures were performed, which resulted in $5563$ data points. The total number of training points after seven cycles of training was 32690. We stopped the training process in the seventh step. The reason for slower growing of data points for the In-Cu-Se system compared with the Ti-O system is the smaller number of seed structures for minima hopping.

\begin{figure}[t]
	\centering
		\includegraphics[scale=.90]{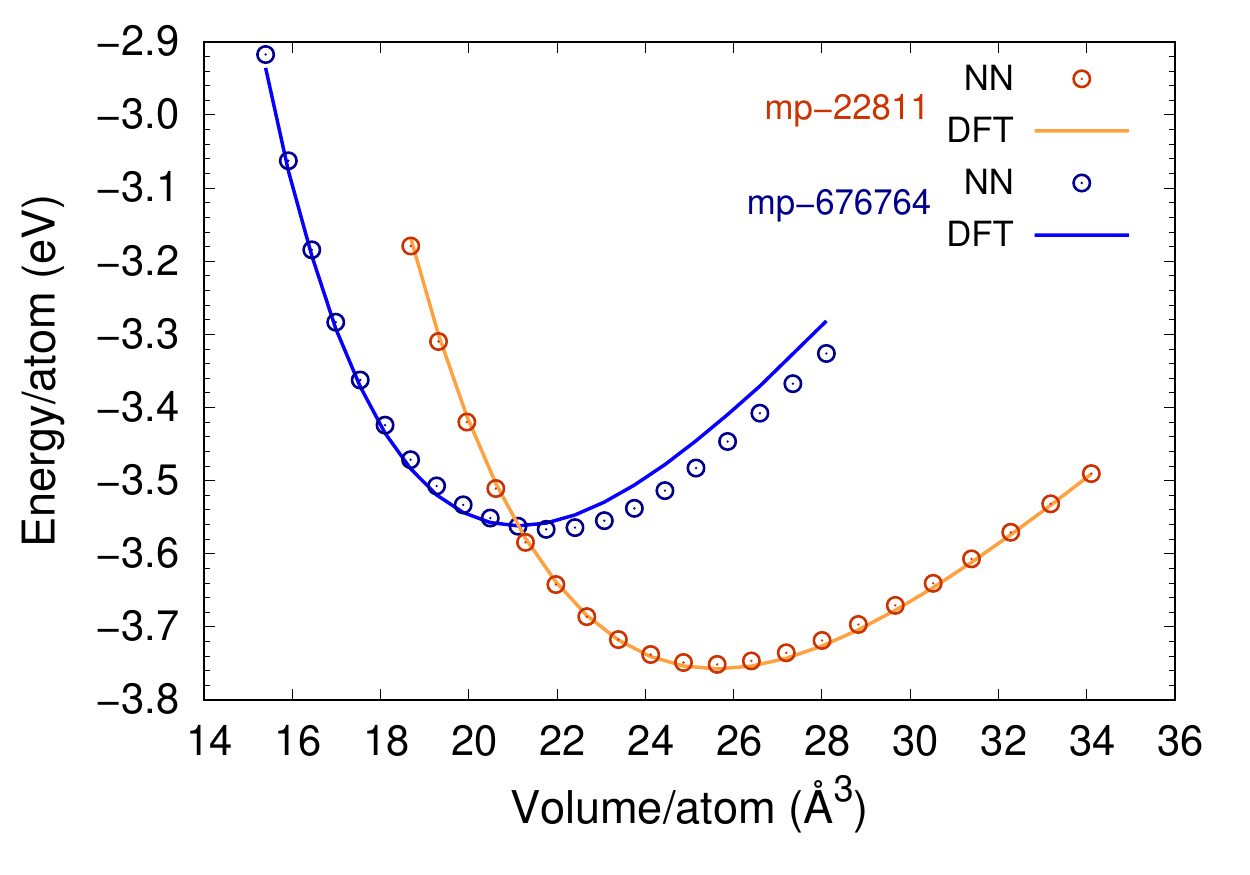}
	\caption{Comparison of the equation of states of two structures of CuInSe$_{\rm 2}$ calculated by DFT and the NNP. The structures are labeled by their Materials Project ID. }
	\label{FIG:CuInSe2_EOS}
\end{figure}

\begin{figure}[t]
	\centering
		\includegraphics[scale=.9]{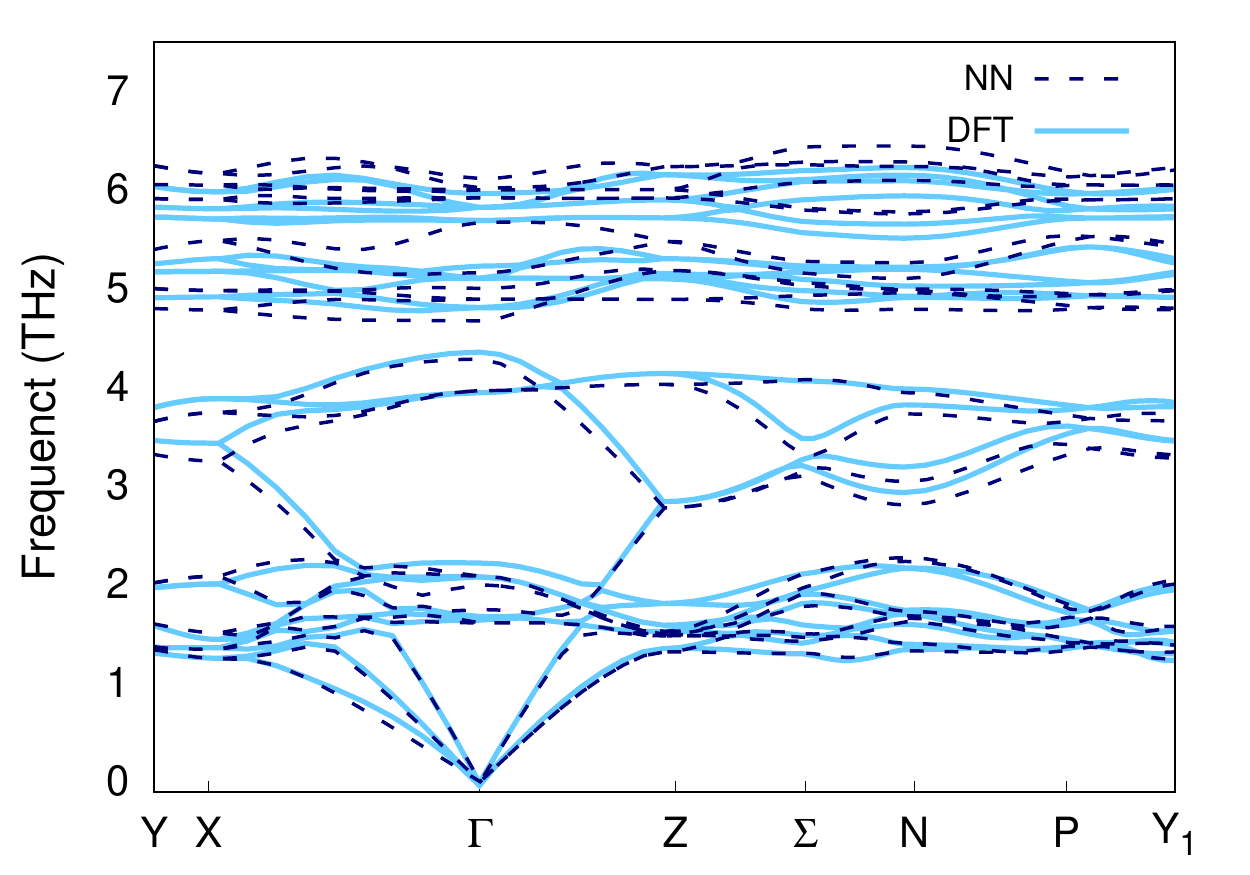}
	\caption{Phonon dispersion relation of chalcopyrite CuInSe$_{\rm 2}$ computed by DFT and the NNP.}
	\label{FIG:CuInSe2_phonon}
\end{figure}

\begin{figure}[t]
	\centering
	\includegraphics[scale=.90]{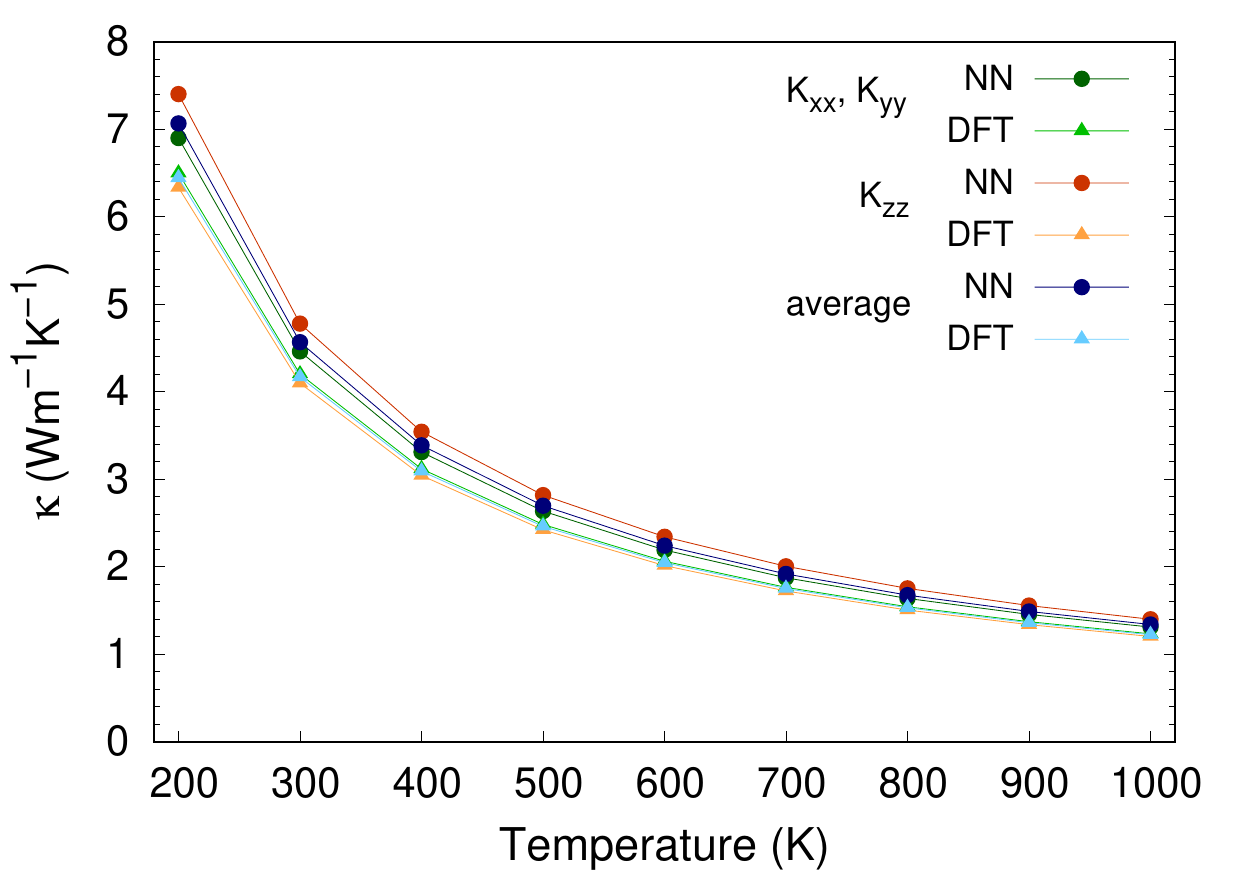}
	\caption{Lattice thermal conductivity of chalcopyrite CuInSe$_{\rm 2}$ calculated with DFT and the NNP.}
	\label{Fig:CuInSe2_kappa}
\end{figure}

Comparing the EOS for two phases of CuInSe$_{\rm 2}$ (Figure~\ref{FIG:CuInSe2_EOS}) confirms the good agreement between DFT and NNP results. The most stable structure of CuInSe$_{\rm 2}$ is chalcopyrite, and the lattice parameters of this structure predicted by DFT are very close to those predicted by the NNP. For the less stable structure of CuInSe$_{\rm 2}$, the lattice parameters predicted by the trained NNP are about $1$\% larger than those computed by DFT. It is noted that the difference between the lattice parameters of this structure calculated with the local density approximation functional and that with the PBE is about 2.5\%. The phonon dispersion relation for chalcopyrite CuInSe$_{\rm 2}$ was calculated for a $3\times3\times3$ supercell and is shown in Figure~\ref{FIG:CuInSe2_phonon}. It is expected that the good agreement between phonon dispersion relations calculated by DFT and by NNP results in a good agreement between calculated values for the lattice thermal conductivity (see Figure~\ref{Fig:CuInSe2_kappa}).

\begin{figure}[ht]
\centering
\includegraphics[scale=.9]{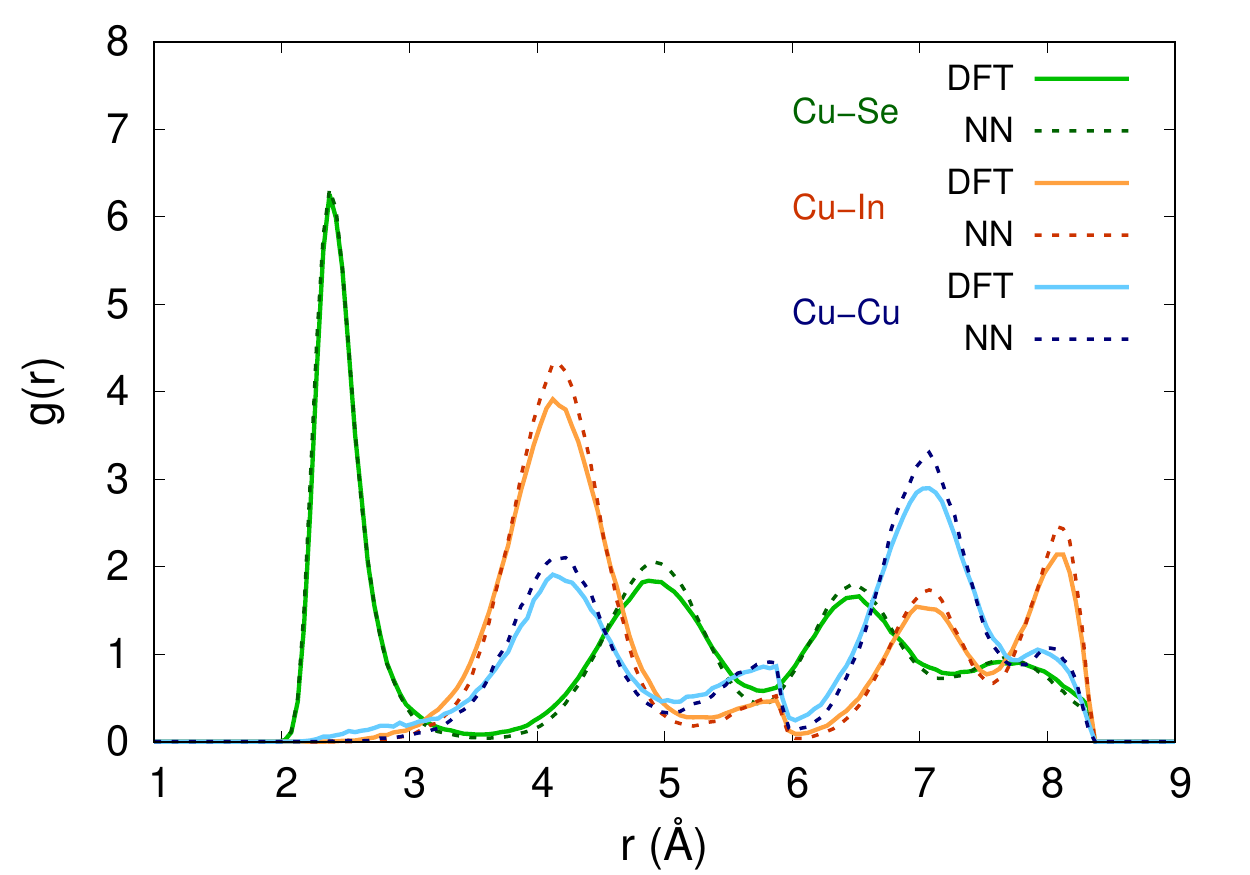}
\caption{The radial distribution functions for the chalcopyrite CuInSe$_{\rm 2}$ supercell computed with DFT and the NNP at T= 1000 K.}
\label{FIG:CuInSe2_VASP-vs-FLAME-1000K}
\end{figure}

To evaluate the accuracy of the constructed NNP for MD simulations, MD calculations at various temperatures were carried out. Figure~\ref{FIG:CuInSe2_VASP-vs-FLAME-1000K} shows the comparison between the radial distribution functions of a 64-atom supercell for a NVT ensemble at $1000$~K for $35000$ MD steps after equilibration. Additionally, we used the climbing image nudged elastic band~\cite{Henkelman2000_2} method to calculate the energy barriers for some known diffusion mechanisms in a supercell of CuInSe$_{\rm 2}$. Saddle points computed with \emph{ab initio} and the NNP are listed in Table~\ref{tbl:CuInSe2_diffusion_barrieres}. Jumping of a Cu atom from a Cu lattice to a Cu vacancy (V$_{\rm Cu}$) site (V$_{\rm Cu}$ $\rightarrow$ V$^{\prime}_{\rm Cu}$) is the most frequent event for diffusion of Cu atoms~\cite{Ramya2019}. The diffusion barriers calculated for this event with DFT and with the generated NNP are in excellent agreement. Another event observed in this system is jumping of an In$_{\rm Cu}$ antisite atom to the V$_{\rm Cu}$ site. The difference between diffusion barriers calculated by DFT and by the NNP for this event is about 5.5\%. 

\begin{table}
    \centering
    \caption{Comparison between diffusion barriers calculated by DFT and the NNP for CuInSe$_{\rm 2}$.}
    \begin{tabular}{lcc}
    \hline
    Process & \multicolumn{2}{c}{Diffusion barrier (eV)}\\
           & DFT & NN \\
    \hline
V$_{\rm Cu}$ $\rightarrow$ V$^{\prime}_{\rm Cu}$ & 1.15 & 1.15\\
In$_{\rm Cu}$ + V$_{\rm Cu}$ $\rightarrow$ In$^{\prime}_{\rm Cu}$ + V$^{\prime}_{\rm Cu}$   & 1.10 & 1.16 \\
\hline
    \end{tabular}
\label{tbl:CuInSe2_diffusion_barrieres}
\end{table}

Regarding surface calculations, the surface energy of the (001) surface calculated by DFT and by the NNP is 0.04 eV/\AA$^{2}$ (0.65 J/m$^{2}$) and 0.04 eV/A$^{2}$ (0.69 J/m$^{2}$), respectively. It should be mentioned that the (001) surface of  CuInSe$_{\rm 2}$ is a relatively `simple' surface. For more `complicated' surfaces, it is necessary to include structures that resemble the surface structure (for example clusters) in the training data~\cite{Faraji2019}.

\section{Summary}
We introduced a Python script for constructing NN interatomic potentials with minimal human intervention. The potential development is performed with algorithms implemented in FLAME. The only data that needs to be provided by the user is the species of the system of interest. The rest of the training process is performed without human control by taking advantage of open-source codes developed for computational materials science such as atomate, Custodian, and WorkFlow. The script keeps track of its steps, and if a failure occurs, the user can continue the potential developing process from the last successfully completed step. The validity of our approach is shown by comparing results of various molecular statics and molecular dynamics calculations based on DFT and NNPs.

\section{Acknowledgements}
The authors from UPB gratefully acknowledge funding of this project by computing time provided by the Paderborn Center for Parallel Computing (PC$^{2}$).

\bibliography{automated_arxiv}

\providecommand{\latin}[1]{#1}
\makeatletter
\providecommand{\doi}
  {\begingroup\let\do\@makeother\dospecials
  \catcode`\{=1 \catcode`\}=2 \doi@aux}
\providecommand{\doi@aux}[1]{\endgroup\texttt{#1}}
\makeatother
\providecommand*\mcitethebibliography{\thebibliography}
\csname @ifundefined\endcsname{endmcitethebibliography}
  {\let\endmcitethebibliography\endthebibliography}{}
\begin{mcitethebibliography}{75}
\providecommand*\natexlab[1]{#1}
\providecommand*\mciteSetBstSublistMode[1]{}
\providecommand*\mciteSetBstMaxWidthForm[2]{}
\providecommand*\mciteBstWouldAddEndPuncttrue
  {\def\EndOfBibitem{\unskip.}}
\providecommand*\mciteBstWouldAddEndPunctfalse
  {\let\EndOfBibitem\relax}
\providecommand*\mciteSetBstMidEndSepPunct[3]{}
\providecommand*\mciteSetBstSublistLabelBeginEnd[3]{}
\providecommand*\EndOfBibitem{}
\mciteSetBstSublistMode{f}
\mciteSetBstMaxWidthForm{subitem}{(\alph{mcitesubitemcount})}
\mciteSetBstSublistLabelBeginEnd
  {\mcitemaxwidthsubitemform\space}
  {\relax}
  {\relax}

\bibitem[Behler(2011)]{Behler2011}
Behler,~J. Atom-centered symmetry functions for constructing high-dimensional
  neural network potentials. \emph{The Journal of Chemical Physics}
  \textbf{2011}, \emph{134}, 074106\relax
\mciteBstWouldAddEndPuncttrue
\mciteSetBstMidEndSepPunct{\mcitedefaultmidpunct}
{\mcitedefaultendpunct}{\mcitedefaultseppunct}\relax
\EndOfBibitem
\bibitem[Behler and Parrinello(2007)Behler, and Parrinello]{Behler2007}
Behler,~J.; Parrinello,~M. Generalized Neural-Network Representation of
  High-Dimensional Potential-Energy Surfaces. \emph{Phys. Rev. Lett.}
  \textbf{2007}, \emph{98}, 146401\relax
\mciteBstWouldAddEndPuncttrue
\mciteSetBstMidEndSepPunct{\mcitedefaultmidpunct}
{\mcitedefaultendpunct}{\mcitedefaultseppunct}\relax
\EndOfBibitem
\bibitem[Bart\'ok \latin{et~al.}(2010)Bart\'ok, Payne, Kondor, and
  Cs\'anyi]{Bartok2010}
Bart\'ok,~A.~P.; Payne,~M.~C.; Kondor,~R.; Cs\'anyi,~G. Gaussian Approximation
  Potentials: The Accuracy of Quantum Mechanics, without the Electrons.
  \emph{Phys. Rev. Lett.} \textbf{2010}, \emph{104}, 136403\relax
\mciteBstWouldAddEndPuncttrue
\mciteSetBstMidEndSepPunct{\mcitedefaultmidpunct}
{\mcitedefaultendpunct}{\mcitedefaultseppunct}\relax
\EndOfBibitem
\bibitem[Botu and Ramprasad(2015)Botu, and Ramprasad]{Botu2014}
Botu,~V.; Ramprasad,~R. Adaptive machine learning framework to accelerate ab
  initio molecular dynamics. \emph{International Journal of Quantum Chemistry}
  \textbf{2015}, \emph{115}, 1074--1083\relax
\mciteBstWouldAddEndPuncttrue
\mciteSetBstMidEndSepPunct{\mcitedefaultmidpunct}
{\mcitedefaultendpunct}{\mcitedefaultseppunct}\relax
\EndOfBibitem
\bibitem[Thompson \latin{et~al.}(2015)Thompson, Swiler, Trott, Foiles, and
  Tucker]{Thompson2015}
Thompson,~A.; Swiler,~L.; Trott,~C.; Foiles,~S.; Tucker,~G. Spectral neighbor
  analysis method for automated generation of quantum-accurate interatomic
  potentials. \emph{Journal of Computational Physics} \textbf{2015},
  \emph{285}, 316 -- 330\relax
\mciteBstWouldAddEndPuncttrue
\mciteSetBstMidEndSepPunct{\mcitedefaultmidpunct}
{\mcitedefaultendpunct}{\mcitedefaultseppunct}\relax
\EndOfBibitem
\bibitem[Li \latin{et~al.}(2015)Li, Kermode, and De~Vita]{Li2015}
Li,~Z.; Kermode,~J.~R.; De~Vita,~A. Molecular Dynamics with On-the-Fly Machine
  Learning of Quantum-Mechanical Forces. \emph{Phys. Rev. Lett.} \textbf{2015},
  \emph{114}, 096405\relax
\mciteBstWouldAddEndPuncttrue
\mciteSetBstMidEndSepPunct{\mcitedefaultmidpunct}
{\mcitedefaultendpunct}{\mcitedefaultseppunct}\relax
\EndOfBibitem
\bibitem[Hansen \latin{et~al.}(2015)Hansen, Biegler, Ramakrishnan, Pronobis,
  von Lilienfeld, Müller, and Tkatchenko]{Hansen2015}
Hansen,~K.; Biegler,~F.; Ramakrishnan,~R.; Pronobis,~W.; von Lilienfeld,~O.~A.;
  Müller,~K.-R.; Tkatchenko,~A. Machine Learning Predictions of Molecular
  Properties: Accurate Many-Body Potentials and Nonlocality in Chemical Space.
  \emph{The Journal of Physical Chemistry Letters} \textbf{2015}, \emph{6},
  2326--2331\relax
\mciteBstWouldAddEndPuncttrue
\mciteSetBstMidEndSepPunct{\mcitedefaultmidpunct}
{\mcitedefaultendpunct}{\mcitedefaultseppunct}\relax
\EndOfBibitem
\bibitem[Brockherde \latin{et~al.}(2017)Brockherde, Vogt, Li, Tuckerman, Burke,
  and M{\"u}ller]{Brockherde2017}
Brockherde,~F.; Vogt,~L.; Li,~L.; Tuckerman,~M.~E.; Burke,~K.;
  M{\"u}ller,~K.-R. Bypassing the Kohn-Sham equations with machine learning.
  \emph{Nature Communications} \textbf{2017}, \emph{8}, 872\relax
\mciteBstWouldAddEndPuncttrue
\mciteSetBstMidEndSepPunct{\mcitedefaultmidpunct}
{\mcitedefaultendpunct}{\mcitedefaultseppunct}\relax
\EndOfBibitem
\bibitem[Smith \latin{et~al.}(2017)Smith, Isayev, and Roitberg]{Smith2017}
Smith,~J.~S.; Isayev,~O.; Roitberg,~A.~E. ANI-1: an extensible neural network
  potential with DFT accuracy at force field computational cost. \emph{Chem.
  Sci.} \textbf{2017}, \emph{8}, 3192--3203\relax
\mciteBstWouldAddEndPuncttrue
\mciteSetBstMidEndSepPunct{\mcitedefaultmidpunct}
{\mcitedefaultendpunct}{\mcitedefaultseppunct}\relax
\EndOfBibitem
\bibitem[Yao \latin{et~al.}(2018)Yao, Herr, Toth, Mckintyre, and
  Parkhill]{Yao2018}
Yao,~K.; Herr,~J.~E.; Toth,~D.; Mckintyre,~R.; Parkhill,~J. The TensorMol-0.1
  model chemistry: a neural network augmented with long-range physics.
  \emph{Chem. Sci.} \textbf{2018}, \emph{9}, 2261--2269\relax
\mciteBstWouldAddEndPuncttrue
\mciteSetBstMidEndSepPunct{\mcitedefaultmidpunct}
{\mcitedefaultendpunct}{\mcitedefaultseppunct}\relax
\EndOfBibitem
\bibitem[Podryabinkin and Shapeev(2017)Podryabinkin, and
  Shapeev]{Podryabinkin2017}
Podryabinkin,~E.~V.; Shapeev,~A.~V. Active learning of linearly parametrized
  interatomic potentials. \emph{Computational Materials Science} \textbf{2017},
  \emph{140}, 171 -- 180\relax
\mciteBstWouldAddEndPuncttrue
\mciteSetBstMidEndSepPunct{\mcitedefaultmidpunct}
{\mcitedefaultendpunct}{\mcitedefaultseppunct}\relax
\EndOfBibitem
\bibitem[Zhang \latin{et~al.}(2018)Zhang, Han, Wang, Car, and E]{Zhang2018}
Zhang,~L.; Han,~J.; Wang,~H.; Car,~R.; E,~W. Deep Potential Molecular Dynamics:
  A Scalable Model with the Accuracy of Quantum Mechanics. \emph{Phys. Rev.
  Lett.} \textbf{2018}, \emph{120}, 143001\relax
\mciteBstWouldAddEndPuncttrue
\mciteSetBstMidEndSepPunct{\mcitedefaultmidpunct}
{\mcitedefaultendpunct}{\mcitedefaultseppunct}\relax
\EndOfBibitem
\bibitem[Bereau \latin{et~al.}(2018)Bereau, DiStasio, Tkatchenko, and von
  Lilienfeld]{Bereau2018}
Bereau,~T.; DiStasio,~R.~A.; Tkatchenko,~A.; von Lilienfeld,~O.~A. Non-covalent
  interactions across organic and biological subsets of chemical space:
  Physics-based potentials parametrized from machine learning. \emph{The
  Journal of Chemical Physics} \textbf{2018}, \emph{148}, 241706\relax
\mciteBstWouldAddEndPuncttrue
\mciteSetBstMidEndSepPunct{\mcitedefaultmidpunct}
{\mcitedefaultendpunct}{\mcitedefaultseppunct}\relax
\EndOfBibitem
\bibitem[Schütt \latin{et~al.}(2019)Schütt, Kessel, Gastegger, Nicoli,
  Tkatchenko, and Müller]{Schuett2019}
Schütt,~K.~T.; Kessel,~P.; Gastegger,~M.; Nicoli,~K.~A.; Tkatchenko,~A.;
  Müller,~K.-R. SchNetPack: A Deep Learning Toolbox For Atomistic Systems.
  \emph{Journal of Chemical Theory and Computation} \textbf{2019}, \emph{15},
  448--455\relax
\mciteBstWouldAddEndPuncttrue
\mciteSetBstMidEndSepPunct{\mcitedefaultmidpunct}
{\mcitedefaultendpunct}{\mcitedefaultseppunct}\relax
\EndOfBibitem
\bibitem[Khaliullin \latin{et~al.}(2010)Khaliullin, Eshet, K\"uhne, Behler, and
  Parrinello]{PhysRevB.81.100103}
Khaliullin,~R.~Z.; Eshet,~H.; K\"uhne,~T.~D.; Behler,~J.; Parrinello,~M.
  Graphite-diamond phase coexistence study employing a neural-network mapping
  of the ab initio potential energy surface. \emph{Phys. Rev. B} \textbf{2010},
  \emph{81}, 100103\relax
\mciteBstWouldAddEndPuncttrue
\mciteSetBstMidEndSepPunct{\mcitedefaultmidpunct}
{\mcitedefaultendpunct}{\mcitedefaultseppunct}\relax
\EndOfBibitem
\bibitem[Eshet \latin{et~al.}(2010)Eshet, Khaliullin, K\"uhne, Behler, and
  Parrinello]{PhysRevB.81.184107}
Eshet,~H.; Khaliullin,~R.~Z.; K\"uhne,~T.~D.; Behler,~J.; Parrinello,~M. Ab
  initio quality neural-network potential for sodium. \emph{Phys. Rev. B}
  \textbf{2010}, \emph{81}, 184107\relax
\mciteBstWouldAddEndPuncttrue
\mciteSetBstMidEndSepPunct{\mcitedefaultmidpunct}
{\mcitedefaultendpunct}{\mcitedefaultseppunct}\relax
\EndOfBibitem
\bibitem[Artrith \latin{et~al.}(2011)Artrith, Morawietz, and
  Behler]{Artrith2011}
Artrith,~N.; Morawietz,~T.; Behler,~J. High-dimensional neural-network
  potentials for multicomponent systems: Applications to zinc oxide.
  \emph{Phys. Rev. B} \textbf{2011}, \emph{83}, 153101\relax
\mciteBstWouldAddEndPuncttrue
\mciteSetBstMidEndSepPunct{\mcitedefaultmidpunct}
{\mcitedefaultendpunct}{\mcitedefaultseppunct}\relax
\EndOfBibitem
\bibitem[Khaliullin \latin{et~al.}(2011)Khaliullin, Eshet, K\"uhne, Behler, and
  Parrinello]{Khaliullin2011}
Khaliullin,~R.~Z.; Eshet,~H.; K\"uhne,~T.~D.; Behler,~J.; Parrinello,~M.
  Nucleation mechanism for the direct graphite-to-diamond phase transition.
  \emph{Nature Mater.} \textbf{2011}, \emph{10}, 693\relax
\mciteBstWouldAddEndPuncttrue
\mciteSetBstMidEndSepPunct{\mcitedefaultmidpunct}
{\mcitedefaultendpunct}{\mcitedefaultseppunct}\relax
\EndOfBibitem
\bibitem[Eshet \latin{et~al.}(2012)Eshet, Khaliullin, K\"uhne, Behler, and
  Parrinello]{PhysRevLett.108.115701}
Eshet,~H.; Khaliullin,~R.~Z.; K\"uhne,~T.~D.; Behler,~J.; Parrinello,~M.
  Microscopic Origins of the Anomalous Melting Behavior of Sodium under High
  Pressure. \emph{Phys. Rev. Lett.} \textbf{2012}, \emph{108}, 115701\relax
\mciteBstWouldAddEndPuncttrue
\mciteSetBstMidEndSepPunct{\mcitedefaultmidpunct}
{\mcitedefaultendpunct}{\mcitedefaultseppunct}\relax
\EndOfBibitem
\bibitem[Artrith~N.(2016)]{Artrith2016}
Artrith~N.,~U.~A. An implementation of artificial neural-network potentials for
  atomistic materials simulations: Performance for TiO2. \emph{Comput. Mater.
  Sci.} \textbf{2016}, \emph{114}, 135\relax
\mciteBstWouldAddEndPuncttrue
\mciteSetBstMidEndSepPunct{\mcitedefaultmidpunct}
{\mcitedefaultendpunct}{\mcitedefaultseppunct}\relax
\EndOfBibitem
\bibitem[Eckhoff and Behler(2019)Eckhoff, and Behler]{Eckhoff2019}
Eckhoff,~M.; Behler,~J. From Molecular Fragments to the Bulk: Development of a
  Neural Network Potential for MOF-5. \emph{Journal of Chemical Theory and
  Computation} \textbf{2019}, \emph{15}, 3793--3809\relax
\mciteBstWouldAddEndPuncttrue
\mciteSetBstMidEndSepPunct{\mcitedefaultmidpunct}
{\mcitedefaultendpunct}{\mcitedefaultseppunct}\relax
\EndOfBibitem
\bibitem[Morawietz \latin{et~al.}(2016)Morawietz, Singraber, Dellago, and
  Behler]{Morawietz2016}
Morawietz,~T.; Singraber,~A.; Dellago,~C.; Behler,~J. How van der Waals
  interactions determine the unique properties of water. \emph{Proceedings of
  the National Academy of Sciences} \textbf{2016}, \emph{113}, 836--8373\relax
\mciteBstWouldAddEndPuncttrue
\mciteSetBstMidEndSepPunct{\mcitedefaultmidpunct}
{\mcitedefaultendpunct}{\mcitedefaultseppunct}\relax
\EndOfBibitem
\bibitem[Sukuba \latin{et~al.}(2018)Sukuba, Chen, Probst, and
  Kaiser]{Sukuba2018}
Sukuba,~I.; Chen,~L.; Probst,~M.; Kaiser,~A. A neural network interface for
  DL\_POLY and its application to liquid water. \emph{Molecular Simulation}
  \textbf{2018}, \emph{0}, 1--6\relax
\mciteBstWouldAddEndPuncttrue
\mciteSetBstMidEndSepPunct{\mcitedefaultmidpunct}
{\mcitedefaultendpunct}{\mcitedefaultseppunct}\relax
\EndOfBibitem
\bibitem[Natarajan and Behler(2016)Natarajan, and Behler]{Natarajan2016}
Natarajan,~S.~K.; Behler,~J. Neural network molecular dynamics simulations of
  solid–liquid interfaces: water at low-index copper surfaces. \emph{Phys.
  Chem. Chem. Phys.} \textbf{2016}, \emph{18}, 28704--28725\relax
\mciteBstWouldAddEndPuncttrue
\mciteSetBstMidEndSepPunct{\mcitedefaultmidpunct}
{\mcitedefaultendpunct}{\mcitedefaultseppunct}\relax
\EndOfBibitem
\bibitem[Quaranta \latin{et~al.}(2017)Quaranta, Hellström, and
  Behler]{Quaranta2017}
Quaranta,~V.; Hellström,~M.; Behler,~J. Proton-Transfer Mechanisms at the
  Water–ZnO Interface: The Role of Presolvation. \emph{The Journal of
  Physical Chemistry Letters} \textbf{2017}, \emph{8}, 1476--1483, PMID:
  28296415\relax
\mciteBstWouldAddEndPuncttrue
\mciteSetBstMidEndSepPunct{\mcitedefaultmidpunct}
{\mcitedefaultendpunct}{\mcitedefaultseppunct}\relax
\EndOfBibitem
\bibitem[Quaranta \latin{et~al.}(2019)Quaranta, Behler, and
  Hellström]{Quaranta2019}
Quaranta,~V.; Behler,~J.; Hellström,~M. Structure and Dynamics of the
  Liquid–Water/Zinc-Oxide Interface from Machine Learning Potential
  Simulations. \emph{The Journal of Physical Chemistry C} \textbf{2019},
  \emph{123}, 1293--1304\relax
\mciteBstWouldAddEndPuncttrue
\mciteSetBstMidEndSepPunct{\mcitedefaultmidpunct}
{\mcitedefaultendpunct}{\mcitedefaultseppunct}\relax
\EndOfBibitem
\bibitem[Hellström \latin{et~al.}(2019)Hellström, Quaranta, and
  Behler]{Hellstroem2019}
Hellström,~M.; Quaranta,~V.; Behler,~J. One-dimensional vs. two-dimensional
  proton transport processes at solid–liquid zinc-oxide–water interfaces.
  \emph{Chem. Sci.} \textbf{2019}, \emph{10}, 1232--1243\relax
\mciteBstWouldAddEndPuncttrue
\mciteSetBstMidEndSepPunct{\mcitedefaultmidpunct}
{\mcitedefaultendpunct}{\mcitedefaultseppunct}\relax
\EndOfBibitem
\bibitem[Ludwig \latin{et~al.}(2019)Ludwig, Gauthier, Brown, Ringe, Nørskov,
  and Chan]{Ludwig2019}
Ludwig,~T.; Gauthier,~J.~A.; Brown,~K.~S.; Ringe,~S.; Nørskov,~J.~K.; Chan,~K.
  Solvent–Adsorbate Interactions and Adsorbate-Specific Solvent Structure in
  Carbon Dioxide Reduction on a Stepped Cu Surface. \emph{The Journal of
  Physical Chemistry C} \textbf{2019}, \emph{123}, 5999--6009\relax
\mciteBstWouldAddEndPuncttrue
\mciteSetBstMidEndSepPunct{\mcitedefaultmidpunct}
{\mcitedefaultendpunct}{\mcitedefaultseppunct}\relax
\EndOfBibitem
\bibitem[Li \latin{et~al.}(2017)Li, Ando, Minamitani, and Watanabe]{Li2017}
Li,~W.; Ando,~Y.; Minamitani,~E.; Watanabe,~S. Study of Li atom diffusion in
  amorphous Li3PO4 with neural network potential. \emph{The Journal of Chemical
  Physics} \textbf{2017}, \emph{147}, 214106\relax
\mciteBstWouldAddEndPuncttrue
\mciteSetBstMidEndSepPunct{\mcitedefaultmidpunct}
{\mcitedefaultendpunct}{\mcitedefaultseppunct}\relax
\EndOfBibitem
\bibitem[Korolev \latin{et~al.}(2020)Korolev, Mitrofanov, Nevolin, Krotov,
  Ul'yanov, and Protsenko]{Korolev2020}
Korolev,~V.~V.; Mitrofanov,~A.~A.; Nevolin,~Y.~M.; Krotov,~V.~V.;
  Ul'yanov,~D.~K.; Protsenko,~P.~V. Neural Network Based Modeling of Grain
  Boundary Complexions Localized in Simple Symmetric Tilt Boundaries $\Sigma$3
  (111) and $\Sigma$5 (210). \emph{Colloid Journal} \textbf{2020}, \emph{82},
  689--695\relax
\mciteBstWouldAddEndPuncttrue
\mciteSetBstMidEndSepPunct{\mcitedefaultmidpunct}
{\mcitedefaultendpunct}{\mcitedefaultseppunct}\relax
\EndOfBibitem
\bibitem[Elbaz \latin{et~al.}(2020)Elbaz, Furman, and
  Caspary~Toroker]{Elbaz2020}
Elbaz,~Y.; Furman,~D.; Caspary~Toroker,~M. Modeling Diffusion in Functional
  Materials: From Density Functional Theory to Artificial Intelligence.
  \emph{Advanced Functional Materials} \textbf{2020}, \emph{30}, 1900778\relax
\mciteBstWouldAddEndPuncttrue
\mciteSetBstMidEndSepPunct{\mcitedefaultmidpunct}
{\mcitedefaultendpunct}{\mcitedefaultseppunct}\relax
\EndOfBibitem
\bibitem[Xu \latin{et~al.}(2020)Xu, Shi, He, and Shao]{Xu2020}
Xu,~N.; Shi,~Y.; He,~Y.; Shao,~Q. A Deep-Learning Potential for Crystalline and
  Amorphous Li–Si Alloys. \emph{The Journal of Physical Chemistry C}
  \textbf{2020}, \emph{124}, 16278--16288\relax
\mciteBstWouldAddEndPuncttrue
\mciteSetBstMidEndSepPunct{\mcitedefaultmidpunct}
{\mcitedefaultendpunct}{\mcitedefaultseppunct}\relax
\EndOfBibitem
\bibitem[Jain \latin{et~al.}(2013)Jain, Ong, Hautier, Chen, Richards, Dacek,
  Cholia, Gunter, Skinner, Ceder, and Persson]{MPD1}
Jain,~A.; Ong,~S.~P.; Hautier,~G.; Chen,~W.; Richards,~W.~D.; Dacek,~S.;
  Cholia,~S.; Gunter,~D.; Skinner,~D.; Ceder,~G.; Persson,~K.~A. Commentary:
  The Materials Project: A materials genome approach to accelerating materials
  innovation. \emph{APL Materials} \textbf{2013}, \emph{1}, 011002\relax
\mciteBstWouldAddEndPuncttrue
\mciteSetBstMidEndSepPunct{\mcitedefaultmidpunct}
{\mcitedefaultendpunct}{\mcitedefaultseppunct}\relax
\EndOfBibitem
\bibitem[Ong \latin{et~al.}(2015)Ong, Cholia, Jain, Brafman, Gunter, Ceder, and
  Persson]{MPD2}
Ong,~S.~P.; Cholia,~S.; Jain,~A.; Brafman,~M.; Gunter,~D.; Ceder,~G.;
  Persson,~K.~A. The Materials Application Programming Interface (API): A
  simple, flexible and efficient API for materials data based on
  REpresentational State Transfer (REST) principles. \emph{Computational
  Materials Science} \textbf{2015}, \emph{97}, 209 -- 215\relax
\mciteBstWouldAddEndPuncttrue
\mciteSetBstMidEndSepPunct{\mcitedefaultmidpunct}
{\mcitedefaultendpunct}{\mcitedefaultseppunct}\relax
\EndOfBibitem
\bibitem[Hautier \latin{et~al.}(2011)Hautier, Fischer, Ehrlacher, Jain, and
  Ceder]{Hautier2011}
Hautier,~G.; Fischer,~C.; Ehrlacher,~V.; Jain,~A.; Ceder,~G. Data Mined Ionic
  Substitutions for the Discovery of New Compounds. \emph{Inorganic Chemistry}
  \textbf{2011}, \emph{50}, 656--663, PMID: 21142147\relax
\mciteBstWouldAddEndPuncttrue
\mciteSetBstMidEndSepPunct{\mcitedefaultmidpunct}
{\mcitedefaultendpunct}{\mcitedefaultseppunct}\relax
\EndOfBibitem
\bibitem[Goedecker(2004)]{Goedecker2004}
Goedecker,~S. Minima hopping: An efficient search method for the global minimum
  of the potential energy surface of complex molecular systems. \emph{The
  Journal of Chemical Physics} \textbf{2004}, \emph{120}, 9911--9917\relax
\mciteBstWouldAddEndPuncttrue
\mciteSetBstMidEndSepPunct{\mcitedefaultmidpunct}
{\mcitedefaultendpunct}{\mcitedefaultseppunct}\relax
\EndOfBibitem
\bibitem[Amsler and Goedecker(2010)Amsler, and Goedecker]{Amsler2010}
Amsler,~M.; Goedecker,~S. Crystal structure prediction using the minima hopping
  method. \emph{J. Chem. Phys.} \textbf{2010}, \emph{133}, 224104\relax
\mciteBstWouldAddEndPuncttrue
\mciteSetBstMidEndSepPunct{\mcitedefaultmidpunct}
{\mcitedefaultendpunct}{\mcitedefaultseppunct}\relax
\EndOfBibitem
\bibitem[Kresse and Hafner(1993)Kresse, and Hafner]{Kresse1993}
Kresse,~G.; Hafner,~J. Ab initio molecular dynamics for liquid metals.
  \emph{Phys. Rev. B} \textbf{1993}, \emph{47}, 558--561\relax
\mciteBstWouldAddEndPuncttrue
\mciteSetBstMidEndSepPunct{\mcitedefaultmidpunct}
{\mcitedefaultendpunct}{\mcitedefaultseppunct}\relax
\EndOfBibitem
\bibitem[Kresse and Furthm\"uller(1996)Kresse, and Furthm\"uller]{Kresse1996}
Kresse,~G.; Furthm\"uller,~J. Efficient iterative schemes for ab initio
  total-energy calculations using a plane-wave basis set. \emph{Phys. Rev. B}
  \textbf{1996}, \emph{54}, 11169--11186\relax
\mciteBstWouldAddEndPuncttrue
\mciteSetBstMidEndSepPunct{\mcitedefaultmidpunct}
{\mcitedefaultendpunct}{\mcitedefaultseppunct}\relax
\EndOfBibitem
\bibitem[Kresse and Furthmüller(1996)Kresse, and Furthmüller]{Kresse1996a}
Kresse,~G.; Furthmüller,~J. Efficiency of ab-initio total energy calculations
  for metals and semiconductors using a plane-wave basis set.
  \emph{Computational Materials Science} \textbf{1996}, \emph{6}, 15 --
  50\relax
\mciteBstWouldAddEndPuncttrue
\mciteSetBstMidEndSepPunct{\mcitedefaultmidpunct}
{\mcitedefaultendpunct}{\mcitedefaultseppunct}\relax
\EndOfBibitem
\bibitem[Bl\"ochl(1994)]{bloechl_1994}
Bl\"ochl,~P.~E. Projector augmented-wave method. \emph{Phys. Rev. B}
  \textbf{1994}, \emph{50}, 17953--17979\relax
\mciteBstWouldAddEndPuncttrue
\mciteSetBstMidEndSepPunct{\mcitedefaultmidpunct}
{\mcitedefaultendpunct}{\mcitedefaultseppunct}\relax
\EndOfBibitem
\bibitem[Perdew \latin{et~al.}(1996)Perdew, Burke, and Ernzerhof]{Perdew1996}
Perdew,~J.~P.; Burke,~K.; Ernzerhof,~M. Generalized Gradient Approximation Made
  Simple. \emph{Phys. Rev. Lett.} \textbf{1996}, \emph{77}, 3865--3868\relax
\mciteBstWouldAddEndPuncttrue
\mciteSetBstMidEndSepPunct{\mcitedefaultmidpunct}
{\mcitedefaultendpunct}{\mcitedefaultseppunct}\relax
\EndOfBibitem
\bibitem[Togo and Tanaka(2015)Togo, and Tanaka]{Togo2015}
Togo,~A.; Tanaka,~I. First principles phonon calculations in materials science.
  \emph{Scripta Materialia} \textbf{2015}, \emph{108}, 1--5\relax
\mciteBstWouldAddEndPuncttrue
\mciteSetBstMidEndSepPunct{\mcitedefaultmidpunct}
{\mcitedefaultendpunct}{\mcitedefaultseppunct}\relax
\EndOfBibitem
\bibitem[Li \latin{et~al.}(2014)Li, Carrete, A.~Katcho, and Mingo]{Li2014Jun}
Li,~W.; Carrete,~J.; A.~Katcho,~N.; Mingo,~N. {ShengBTE: A solver of the
  Boltzmann transport equation for phonons}. \emph{Comput. Phys. Commun.}
  \textbf{2014}, \emph{185}, 1747--1758\relax
\mciteBstWouldAddEndPuncttrue
\mciteSetBstMidEndSepPunct{\mcitedefaultmidpunct}
{\mcitedefaultendpunct}{\mcitedefaultseppunct}\relax
\EndOfBibitem
\bibitem[Mathew \latin{et~al.}(2017)Mathew, Montoya, Faghaninia, Dwarakanath,
  Aykol, Tang, heng Chu, Smidt, Bocklund, Horton, Dagdelen, Wood, Liu, Neaton,
  Ong, Persson, and Jain]{atomate}
Mathew,~K. \latin{et~al.}  Atomate: A high-level interface to generate,
  execute, and analyze computational materials science workflows.
  \emph{Computational Materials Science} \textbf{2017}, \emph{139}, 140 --
  152\relax
\mciteBstWouldAddEndPuncttrue
\mciteSetBstMidEndSepPunct{\mcitedefaultmidpunct}
{\mcitedefaultendpunct}{\mcitedefaultseppunct}\relax
\EndOfBibitem
\bibitem[Ong \latin{et~al.}(2013)Ong, Richards, Jain, Hautier, Kocher, Cholia,
  Gunter, Chevrier, Persson, and Ceder]{pymatgen}
Ong,~S.~P.; Richards,~W.~D.; Jain,~A.; Hautier,~G.; Kocher,~M.; Cholia,~S.;
  Gunter,~D.; Chevrier,~V.~L.; Persson,~K.~A.; Ceder,~G. Python Materials
  Genomics (pymatgen): A robust, open-source python library for materials
  analysis. \emph{Computational Materials Science} \textbf{2013}, \emph{68},
  314 -- 319\relax
\mciteBstWouldAddEndPuncttrue
\mciteSetBstMidEndSepPunct{\mcitedefaultmidpunct}
{\mcitedefaultendpunct}{\mcitedefaultseppunct}\relax
\EndOfBibitem
\bibitem[Jain \latin{et~al.}(2015)Jain, Ong, Chen, Medasani, Qu, Kocher,
  Brafman, Petretto, Rignanese, Hautier, Gunter, and Persson]{Jain2015}
Jain,~A.; Ong,~S.~P.; Chen,~W.; Medasani,~B.; Qu,~X.; Kocher,~M.; Brafman,~M.;
  Petretto,~G.; Rignanese,~G.-M.; Hautier,~G.; Gunter,~D.; Persson,~K.~A.
  FireWorks: a dynamic workflow system designed for high-throughput
  applications. \emph{Concurrency and Computation: Practice and Experience}
  \textbf{2015}, \emph{27}, 5037--5059, CPE-14-0307.R2\relax
\mciteBstWouldAddEndPuncttrue
\mciteSetBstMidEndSepPunct{\mcitedefaultmidpunct}
{\mcitedefaultendpunct}{\mcitedefaultseppunct}\relax
\EndOfBibitem
\bibitem[Amsler \latin{et~al.}(2020)Amsler, Rostami, Tahmasbi, Khajehpasha,
  Faraji, Rasoulkhani, and Ghasemi]{Amsler2020}
Amsler,~M.; Rostami,~S.; Tahmasbi,~H.; Khajehpasha,~E.~R.; Faraji,~S.;
  Rasoulkhani,~R.; Ghasemi,~S.~A. FLAME: A library of atomistic modeling
  environments. \emph{Computer Physics Communications} \textbf{2020},
  \emph{256}, 107415\relax
\mciteBstWouldAddEndPuncttrue
\mciteSetBstMidEndSepPunct{\mcitedefaultmidpunct}
{\mcitedefaultendpunct}{\mcitedefaultseppunct}\relax
\EndOfBibitem
\bibitem[fla(2018)]{flame}
FLAME: a library of atomistic modeling environments. 2018;
  \url{https://flame-code.org} and
  \url{https://github.com/flame-code/FLAME}\relax
\mciteBstWouldAddEndPuncttrue
\mciteSetBstMidEndSepPunct{\mcitedefaultmidpunct}
{\mcitedefaultendpunct}{\mcitedefaultseppunct}\relax
\EndOfBibitem
\bibitem[Ghasemi \latin{et~al.}(2015)Ghasemi, Hofstetter, Saha, and
  Goedecker]{Ghasemi2015}
Ghasemi,~S.~A.; Hofstetter,~A.; Saha,~S.; Goedecker,~S. Interatomic potentials
  for ionic systems with density functional accuracy based on charge densities
  obtained by a neural network. \emph{Phys. Rev. B} \textbf{2015}, \emph{92},
  045131\relax
\mciteBstWouldAddEndPuncttrue
\mciteSetBstMidEndSepPunct{\mcitedefaultmidpunct}
{\mcitedefaultendpunct}{\mcitedefaultseppunct}\relax
\EndOfBibitem
\bibitem[Rostami \latin{et~al.}(2018)Rostami, Amsler, and Ghasemi]{Rostami2018}
Rostami,~S.; Amsler,~M.; Ghasemi,~S.~A. Optimized symmetry functions for
  machine-learning interatomic potentials of multicomponent systems. \emph{J.
  Chem. Phys.} \textbf{2018}, \emph{149}, 124106\relax
\mciteBstWouldAddEndPuncttrue
\mciteSetBstMidEndSepPunct{\mcitedefaultmidpunct}
{\mcitedefaultendpunct}{\mcitedefaultseppunct}\relax
\EndOfBibitem
\bibitem[Faraji \latin{et~al.}(2017)Faraji, Ghasemi, Rostami, Rasoulkhani,
  Schaefer, Goedecker, and Amsler]{Faraji2017}
Faraji,~S.; Ghasemi,~S.~A.; Rostami,~S.; Rasoulkhani,~R.; Schaefer,~B.;
  Goedecker,~S.; Amsler,~M. High accuracy and transferability of a neural
  network potential through charge equilibration for calcium fluoride.
  \emph{Phys. Rev. B} \textbf{2017}, \emph{95}, 104105\relax
\mciteBstWouldAddEndPuncttrue
\mciteSetBstMidEndSepPunct{\mcitedefaultmidpunct}
{\mcitedefaultendpunct}{\mcitedefaultseppunct}\relax
\EndOfBibitem
\bibitem[Rasoulkhani \latin{et~al.}(2017)Rasoulkhani, Tahmasbi, Ghasemi,
  Faraji, Rostami, and Amsler]{Rasoulkhani2017}
Rasoulkhani,~R.; Tahmasbi,~H.; Ghasemi,~S.~A.; Faraji,~S.; Rostami,~S.;
  Amsler,~M. Energy landscape of ZnO clusters and low-density polymorphs.
  \emph{Phys. Rev. B} \textbf{2017}, \emph{96}, 064108\relax
\mciteBstWouldAddEndPuncttrue
\mciteSetBstMidEndSepPunct{\mcitedefaultmidpunct}
{\mcitedefaultendpunct}{\mcitedefaultseppunct}\relax
\EndOfBibitem
\bibitem[Eivari \latin{et~al.}(2017)Eivari, Ghasemi, Tahmasbi, Rostami, Faraji,
  Rasoulkhani, Goedecker, and Amsler]{Asna2017}
Eivari,~H.~A.; Ghasemi,~S.~A.; Tahmasbi,~H.; Rostami,~S.; Faraji,~S.;
  Rasoulkhani,~R.; Goedecker,~S.; Amsler,~M. Two-Dimensional Hexagonal Sheet of
  TiO2. \emph{Chem. Mater.} \textbf{2017}, \emph{29}, 8594\relax
\mciteBstWouldAddEndPuncttrue
\mciteSetBstMidEndSepPunct{\mcitedefaultmidpunct}
{\mcitedefaultendpunct}{\mcitedefaultseppunct}\relax
\EndOfBibitem
\bibitem[Faraji \latin{et~al.}(2019)Faraji, Ghasemi, Parsaeifard, and
  Goedecker]{Faraji2019}
Faraji,~S.; Ghasemi,~S.~A.; Parsaeifard,~B.; Goedecker,~S. Surface
  reconstructions and premelting of the (100) CaF2 surface. \emph{Phys. Chem.
  Chem. Phys.} \textbf{2019}, \emph{21}, 16270\relax
\mciteBstWouldAddEndPuncttrue
\mciteSetBstMidEndSepPunct{\mcitedefaultmidpunct}
{\mcitedefaultendpunct}{\mcitedefaultseppunct}\relax
\EndOfBibitem
\bibitem[Oganov and Valle(2009)Oganov, and Valle]{Oganov2009}
Oganov,~A.~R.; Valle,~M. How to quantify energy landscapes of solids. \emph{The
  Journal of Chemical Physics} \textbf{2009}, \emph{130}, 104504\relax
\mciteBstWouldAddEndPuncttrue
\mciteSetBstMidEndSepPunct{\mcitedefaultmidpunct}
{\mcitedefaultendpunct}{\mcitedefaultseppunct}\relax
\EndOfBibitem
\bibitem[YAM()]{YAML}
The Official YAML Web Site. \url{https://yaml.org/}\relax
\mciteBstWouldAddEndPuncttrue
\mciteSetBstMidEndSepPunct{\mcitedefaultmidpunct}
{\mcitedefaultendpunct}{\mcitedefaultseppunct}\relax
\EndOfBibitem
\bibitem[Mangold \latin{et~al.}(2020)Mangold, Chen, Barbalinardo, Behler,
  Pochet, Termentzidis, Han, Chaput, Lacroix, and Donadio]{Mangold2020}
Mangold,~C.; Chen,~S.; Barbalinardo,~G.; Behler,~J.; Pochet,~P.;
  Termentzidis,~K.; Han,~Y.; Chaput,~L.; Lacroix,~D.; Donadio,~D.
  Transferability of neural network potentials for varying stoichiometry:
  Phonons and thermal conductivity of MnxGey compounds. \emph{Journal of
  Applied Physics} \textbf{2020}, \emph{127}, 244901\relax
\mciteBstWouldAddEndPuncttrue
\mciteSetBstMidEndSepPunct{\mcitedefaultmidpunct}
{\mcitedefaultendpunct}{\mcitedefaultseppunct}\relax
\EndOfBibitem
\bibitem[Benoit \latin{et~al.}(2021)Benoit, Amodeo, Combettes, Khaled, Roux,
  and Lam]{Benoit2021}
Benoit,~M.; Amodeo,~J.; Combettes,~S.; Khaled,~I.; Roux,~A.; Lam,~J. Measuring
  transferability issues in machine-learning force fields: the example of
  gold{\textendash}iron interactions with linearized potentials. \emph{Machine
  Learning: Science and Technology} \textbf{2021}, \emph{2}, 025003\relax
\mciteBstWouldAddEndPuncttrue
\mciteSetBstMidEndSepPunct{\mcitedefaultmidpunct}
{\mcitedefaultendpunct}{\mcitedefaultseppunct}\relax
\EndOfBibitem
\bibitem[FUJISHIMA and HONDA(1972)FUJISHIMA, and HONDA]{Fujishima1972}
FUJISHIMA,~A.; HONDA,~K. Electrochemical Photolysis of Water at a Semiconductor
  Electrode. \emph{Nature} \textbf{1972}, \emph{238}, 37--38\relax
\mciteBstWouldAddEndPuncttrue
\mciteSetBstMidEndSepPunct{\mcitedefaultmidpunct}
{\mcitedefaultendpunct}{\mcitedefaultseppunct}\relax
\EndOfBibitem
\bibitem[Ma \latin{et~al.}(2014)Ma, Wang, Jia, Chen, Han, and Li]{Ma2014}
Ma,~Y.; Wang,~X.; Jia,~Y.; Chen,~X.; Han,~H.; Li,~C. Titanium Dioxide-Based
  Nanomaterials for Photocatalytic Fuel Generations. \emph{Chemical Reviews}
  \textbf{2014}, \emph{114}, 9987--10043\relax
\mciteBstWouldAddEndPuncttrue
\mciteSetBstMidEndSepPunct{\mcitedefaultmidpunct}
{\mcitedefaultendpunct}{\mcitedefaultseppunct}\relax
\EndOfBibitem
\bibitem[Ni \latin{et~al.}(2007)Ni, Leung, Leung, and Sumathy]{Ni2007}
Ni,~M.; Leung,~M.~K.; Leung,~D.~Y.; Sumathy,~K. A review and recent
  developments in photocatalytic water-splitting using TiO2 for hydrogen
  production. \emph{Renewable and Sustainable Energy Reviews} \textbf{2007},
  \emph{11}, 401 -- 425\relax
\mciteBstWouldAddEndPuncttrue
\mciteSetBstMidEndSepPunct{\mcitedefaultmidpunct}
{\mcitedefaultendpunct}{\mcitedefaultseppunct}\relax
\EndOfBibitem
\bibitem[Slack(1973)]{Slack1973}
Slack,~G. Nonmetallic crystals with high thermal conductivity. \emph{Journal of
  Physics and Chemistry of Solids} \textbf{1973}, \emph{34}, 321--335\relax
\mciteBstWouldAddEndPuncttrue
\mciteSetBstMidEndSepPunct{\mcitedefaultmidpunct}
{\mcitedefaultendpunct}{\mcitedefaultseppunct}\relax
\EndOfBibitem
\bibitem[Shojaee and Mohammadizadeh(2009)Shojaee, and
  Mohammadizadeh]{Shojaee2009}
Shojaee,~E.; Mohammadizadeh,~M.~R. First-principles elastic and thermal
  properties of {TiO}2: a phonon approach. \emph{Journal of Physics: Condensed
  Matter} \textbf{2009}, \emph{22}, 015401\relax
\mciteBstWouldAddEndPuncttrue
\mciteSetBstMidEndSepPunct{\mcitedefaultmidpunct}
{\mcitedefaultendpunct}{\mcitedefaultseppunct}\relax
\EndOfBibitem
\bibitem[Arrigoni and Madsen(2019)Arrigoni, and Madsen]{Arrigoni2019}
Arrigoni,~M.; Madsen,~G.~K. Comparing the performance of LDA and GGA
  functionals in predicting the lattice thermal conductivity of III-V
  semiconductor materials in the zincblende structure: The cases of AlAs and
  BAs. \emph{Computational Materials Science} \textbf{2019}, \emph{156}, 354 --
  360\relax
\mciteBstWouldAddEndPuncttrue
\mciteSetBstMidEndSepPunct{\mcitedefaultmidpunct}
{\mcitedefaultendpunct}{\mcitedefaultseppunct}\relax
\EndOfBibitem
\bibitem[Torres and Rurali(2019)Torres, and Rurali]{Torres2019}
Torres,~P.; Rurali,~R. Thermal Conductivity of Rutile and Anatase TiO2 from
  First-Principles. \emph{The Journal of Physical Chemistry C} \textbf{2019},
  \emph{123}, 30851--30855\relax
\mciteBstWouldAddEndPuncttrue
\mciteSetBstMidEndSepPunct{\mcitedefaultmidpunct}
{\mcitedefaultendpunct}{\mcitedefaultseppunct}\relax
\EndOfBibitem
\bibitem[EON()]{EON}
EON: Long timescale dynamics.
  \url{https://theory.cm.utexas.edu/eon/index.html}\relax
\mciteBstWouldAddEndPuncttrue
\mciteSetBstMidEndSepPunct{\mcitedefaultmidpunct}
{\mcitedefaultendpunct}{\mcitedefaultseppunct}\relax
\EndOfBibitem
\bibitem[Fro(2019)]{Frontier2019}
Solar Frontier, Solar Frontier Achieves World Record Thin-Film Solar Cell
  Efficiency of 23.35\%. 2019;
  \url{http://www.solar-frontiercom/eng/news/2019/0117_press.html}, Solar
  Frontier KK. Press release 17.01.2019\relax
\mciteBstWouldAddEndPuncttrue
\mciteSetBstMidEndSepPunct{\mcitedefaultmidpunct}
{\mcitedefaultendpunct}{\mcitedefaultseppunct}\relax
\EndOfBibitem
\bibitem[Regmi \latin{et~al.}(2020)Regmi, Ashok, Chawla, Semalti, Velumani,
  Sharma, and Castaneda]{Regmi2020}
Regmi,~G.; Ashok,~A.; Chawla,~P.; Semalti,~P.; Velumani,~S.; Sharma,~S.~N.;
  Castaneda,~H. Perspectives of chalcopyrite-based CIGSe thin-film solar cell:
  a review. \emph{Journal of Materials Science: Materials in Electronics}
  \textbf{2020}, \emph{31}, 7286--7314\relax
\mciteBstWouldAddEndPuncttrue
\mciteSetBstMidEndSepPunct{\mcitedefaultmidpunct}
{\mcitedefaultendpunct}{\mcitedefaultseppunct}\relax
\EndOfBibitem
\bibitem[Kim \latin{et~al.}(2019)Kim, Park, Hwang, Won, Kim, Lee, and
  Min]{Kim2019}
Kim,~B.; Park,~G.~S.; Hwang,~Y.~J.; Won,~D.~H.; Kim,~W.; Lee,~D.~K.; Min,~B.~K.
  Cu(In,Ga)(S,Se)2 Photocathodes with a Grown-In CuxS Catalyst for Solar Water
  Splitting. \emph{ACS Energy Letters} \textbf{2019}, \emph{4},
  2937--2944\relax
\mciteBstWouldAddEndPuncttrue
\mciteSetBstMidEndSepPunct{\mcitedefaultmidpunct}
{\mcitedefaultendpunct}{\mcitedefaultseppunct}\relax
\EndOfBibitem
\bibitem[Hu \latin{et~al.}(2020)Hu, Gong, Ye, Liu, Xiao, and Yu]{Hu2020}
Hu,~Z.; Gong,~J.; Ye,~Z.; Liu,~Y.; Xiao,~X.; Yu,~J.~C. Cu(In,Ga)Se2 for
  selective and efficient photoelectrochemical conversion of CO2 into CO.
  \emph{Journal of Catalysis} \textbf{2020}, \emph{384}, 88 -- 95\relax
\mciteBstWouldAddEndPuncttrue
\mciteSetBstMidEndSepPunct{\mcitedefaultmidpunct}
{\mcitedefaultendpunct}{\mcitedefaultseppunct}\relax
\EndOfBibitem
\bibitem[Mirhosseini \latin{et~al.}(2020)Mirhosseini, Kormath Madam~Raghupathy,
  Sahoo, Wiebeler, Chugh, and Kühne]{Mirhosseini2020}
Mirhosseini,~H.; Kormath Madam~Raghupathy,~R.; Sahoo,~S.~K.; Wiebeler,~H.;
  Chugh,~M.; Kühne,~T.~D. In silico investigation of Cu(In{,}Ga)Se2-based
  solar cells. \emph{Phys. Chem. Chem. Phys.} \textbf{2020}, \emph{22},
  26682--26701\relax
\mciteBstWouldAddEndPuncttrue
\mciteSetBstMidEndSepPunct{\mcitedefaultmidpunct}
{\mcitedefaultendpunct}{\mcitedefaultseppunct}\relax
\EndOfBibitem
\bibitem[Henkelman \latin{et~al.}(2000)Henkelman, Uberuaga, and
  Jónsson]{Henkelman2000_2}
Henkelman,~G.; Uberuaga,~B.~P.; Jónsson,~H. A climbing image nudged elastic
  band method for finding saddle points and minimum energy paths. \emph{The
  Journal of Chemical Physics} \textbf{2000}, \emph{113}, 9901--9904\relax
\mciteBstWouldAddEndPuncttrue
\mciteSetBstMidEndSepPunct{\mcitedefaultmidpunct}
{\mcitedefaultendpunct}{\mcitedefaultseppunct}\relax
\EndOfBibitem
\bibitem[Kormath Madam~Raghupathy \latin{et~al.}(2019)Kormath Madam~Raghupathy,
  Kühne, Henkelman, and Mirhosseini]{Ramya2019}
Kormath Madam~Raghupathy,~R.; Kühne,~T.~D.; Henkelman,~G.; Mirhosseini,~H.
  Alkali Atoms Diffusion Mechanism in CuInSe2 Explained by Kinetic Monte Carlo
  Simulations. \emph{Advanced Theory and Simulations} \textbf{2019}, \emph{2},
  1900036\relax
\mciteBstWouldAddEndPuncttrue
\mciteSetBstMidEndSepPunct{\mcitedefaultmidpunct}
{\mcitedefaultendpunct}{\mcitedefaultseppunct}\relax
\EndOfBibitem
\end{mcitethebibliography}

\end{document}